\documentclass{aa}
\usepackage{txfonts}

\usepackage{CJKutf8}        
\usepackage{nicefrac}  
\usepackage{siunitx}  
\usepackage{algorithm2e}  
\usepackage{lstautogobble}  

\usepackage{placeins}  
\usepackage{graphicx}  
\usepackage{natbib}  
\usepackage{hyperref}  
\usepackage[nameinlink]{cleveref}  
\usepackage{subcaption}  

\bibpunct{(}{)}{;}{a}{}{,} 

\DeclareMathOperator*{\argmin}{arg\,min}

\SetKw{KwInIterable}{in}
\makeatletter
\g@addto@macro{\@algocf@init}{\SetKwInOut{Parameter}{Parameters}}
\makeatother

\begin{document}

\title{A parsec-scale Galactic 3D dust map out to 1.25 kpc from the Sun}  

\author{Gordian~Edenhofer\inst{1,2,3}\fnmsep\thanks{\email{edh@mpa-garching.mpg.de}}
   \and
   Catherine~Zucker\inst{3,4}
   \and
   Philipp~Frank\inst{1}
   \and
   Andrew~K.~Saydjari\inst{3}
   \and
   Joshua~S.~Speagle (\mbox{\begin{CJK*}{UTF8}{gbsn}沈佳士\end{CJK*}})\inst{5,6,7,8}
   \and
   Douglas~Finkbeiner\inst{3}
   \and
   Torsten~A.~Enßlin\inst{1,2}
}

\institute{
   Max Planck Institute for Astrophysics,
   Karl-Schwarzschild-Straße 1, 85748 Garching bei München, Germany
   \and
   Ludwig Maximilian University of Munich,
   Geschwister-Scholl-Platz 1, 80539 München, Germany
   \and
   Center for Astrophysics $\vert$ Harvard \& Smithsonian,
   60 Garden St., Cambridge, MA 02138
   \and
   Space Telescope Science Institute,
   3700 San Martin Dr, Baltimore, MD 21218
   \and
   Department of Statistical Sciences,
   University of Toronto, Toronto, ON M5G 1Z5, Canada
   \and
   David A. Dunlap Department of Astronomy \& Astrophysics,
   University of Toronto, Toronto, ON M5S 3H4, Canada
   \and
   Dunlap Institute for Astronomy \& Astrophysics,
   University of Toronto, Toronto, ON M5S 3H4, Canada
   \and
   Data Sciences Institute,
   University of Toronto, Toronto, ON M5G 1Z5, Canada
}

\date{}

\abstract
{
   High-resolution 3D maps of interstellar dust are critical for probing the underlying physics shaping the structure of the interstellar medium, and for foreground correction of astrophysical observations affected by dust.
}
{
   We aim to construct a new 3D map of the spatial distribution of interstellar dust extinction out to a distance of \SI{1.25}{kpc} from the Sun.
}
{
   We leveraged distance and extinction estimates to 54 million nearby stars derived from the \textit{Gaia} BP/RP spectra.
   Using the stellar distance and extinction information, we inferred the spatial distribution of dust extinction.
   We modeled the logarithmic dust extinction with a Gaussian process in a spherical coordinate system via iterative charted refinement and a correlation kernel inferred in previous work.
   In total, our posterior has over 661 million degrees of freedom.
   We probed the posterior distribution using the variational inference method MGVI.
}
{
   Our 3D dust map has an angular resolution of up to ${14'}$ ($N_\text{side}=256$), and we achieve parsec-scale distance resolution, sampling the dust in $516$ logarithmically spaced distance bins spanning \SIrange{69}{1250}{pc}.
   We generated 12 samples from the variational posterior of the 3D dust distribution and release the samples alongside the mean 3D dust map and its corresponding uncertainty.
}
{
   Our map resolves the internal structure of hundreds of molecular clouds in the solar neighborhood and will be broadly useful for studies of star formation, Galactic structure, and young stellar populations.
   It is available for download in a variety of coordinate systems online and can also be queried via the publicly available \texttt{dustmaps} Python package.
}

\keywords{
   interstellar dust
   -- interstellar medium
   -- Milky Way
   -- Gaia
   -- Gaussian processes
   -- Bayesian inference
}

\maketitle

\section{Introduction}
\label{sec:introduction}

Interstellar dust comprises only 1\% of the interstellar medium by mass but absorbs and re-radiates $>30\%$ of starlight at infrared wavelengths \citep{Popescu2002}.
As such, dust plays an outsized role in the evolution of galaxies, catalyzing the formation of molecular hydrogen, shielding complex molecules from the UV radiation field, coupling the magnetic field to interstellar gas, and regulating the overall heating and cooling of the interstellar medium \citep{Draine2011}.

Dust's ability to scatter and absorb starlight is precisely the reason why we can probe it in three spatial dimensions.
It preferentially absorbs shorter wavelengths of a stellar spectrum, thus leading to stars behind dense dust clouds appearing reddened relative to their intrinsic colors.
The amount by which stars behind dust clouds appear reddened allows us to infer the amount of dust extinction between us and the reddened star.
In combination with distance measurements to reddened stars, we can de-project the integrated extinction measurements into a 3D map of differential dust extinction.

\textit{Gaia} has been transformative for the field by providing accurate distance information to more than 1 billion stars, primarily within a few kiloparsecs of the Sun.
Precise distances not only improve our knowledge about a star's position, they also break degeneracies inherent in the modeling of extinction and significantly reduce the extinction uncertainties \citep{Zucker2019}.
Thanks to the large quantity of extinction and distance measurements available in the era of large photometric, astrometric, and spectroscopic surveys, we can now probe the 3D distribution of dust in the Milky Way on parsec scales.

A number of 3D dust maps that combine \textit{Gaia} and vast photometric and spectroscopic surveys already exist.
These maps primarily differ in the way they account for the so-called fingers-of-god effect, or the tendency of dust structures to be smeared out along the line of sight (LOS).
The effect stems from superior constraints on stars' plane-of-sky (POS) positions relative to their LOS distance uncertainties.

Three-dimensional dust maps predominantly fall into two categories, each representing a trade-off between angular resolution and distance resolution: reconstructions on a Cartesian grid and reconstructions on a spherical grid.
Cartesian reconstructions commonly feature less pronounced fingers-of-god but scale poorly with the size of the reconstructed volume.
They either encompass a limited volume of the Galaxy~\citep{Leike2020,Leike2019} at a high resolution or cover a larger volume of the Galaxy at a low resolution~\citep{Vergely2022,Lallement2022,Lallement2019,Lallement2018,Capitanio2017}.
Spherical reconstructions often have a much higher resolution and probe larger volumes of the Galaxy but come with more strongly pronounced fingers-of-god artifacts \citep{Green2019,Green2017,Chen2019}.
Alternative approaches using many small reconstructions \citep{Leike2022}, an analytical approach \citep{Rezaei2022,Rezaei2020,Rezaei2018,Rezaei2017}, or inducing point methods \citep{Dharmawardena2022} have so far been unsuccessful in reconstructing dust at high resolution over large volumes without artifacts.

Physical smoothness priors counterbalance the fingers-of-god effect as finger-like structures are a priori unlikely.
In a Cartesian coordinate system it is comparatively easy to incorporate physical priors into the model, such as the distribution of dust being spatially smooth.
Smoothness priors are often incorporated using Gaussian process (GP) priors.
Sparsities and symmetries in the prior can be exploited to efficiently apply a GP to a regular Cartesian coordinate system.

Spherical coordinate systems break these sparsities and symmetries in the prior but are much better aligned with the desired spacing of voxels along the LOS.
Nearby, voxels can be spaced densely, while at greater distances voxels can be spaced further apart.
Naively using a GP prior is infeasible, and approximations either trade fingers-of-god artifacts for other artifacts~\citep{Leike2022} or are too weak to regularize the reconstructions~\citep{Green2019}.

In this work, we present a 3D dust map that achieves high distance and angular resolution and probes a large volume of the Galaxy, all at a feasible computational cost.
The map uses a new GP prior methodology to incorporate smoothness in a spherical coordinate system, mitigating fingers-of-god artifacts.
With a spherical coordinate system we were able to probe dust beyond \SI{1}{kpc} while still resolving nearby dust clouds at parsec-scale resolution.
In \Cref{sec:data} we present the stellar distance and extinction estimates upon which our map is based.
In \Cref{sec:priors} we present our GP prior methodology for incorporating smoothness in a spherical coordinate system.
\Cref{sec:likelihood} describes how we combine the data with our prior model and how we incorporate the distance uncertainties of stars.
In \Cref{sec:posterior_inference} we describe our inference before recapitulating all approximations of the model and their implications in \Cref{sec:caveats}.
Finally, in \Cref{sec:results} we present the final map and compare it to existing 3D dust maps and 2D observations.

\section{Stellar distance and extinction data}
\label{sec:data}

To construct a 3D dust map, we used the stellar distance and extinction estimates from~\citet{Zhang2023}, which are primarily based on the \textit{Gaia} BP/RP spectra (spectral resolution R $\sim 30-100$).
\citet{Zhang2023} adopted a data-driven approach to forward-model the extinction, distance, and intrinsic parameters of each star given the combination of the \textit{Gaia} BP/RP spectra and infrared photometry from the two micron all sky survey (2MASS) and unWISE, a processed catalog based on the wide-field infrared survey explorer (WISE) \citep{Carasco2021,DeAngeli2022,GaiaCollaboration2022,Montegriffo2022,Schlafly2019,Wright2010,Skrutskie2006}.
The model is trained using a subset of stars with higher resolution spectra ($R \sim 1800$) available from the large sky area multi-object fibre spectroscopic telescope (LAMOST) \citep{Wang2022,Xiang2022}.
The resulting catalog contains distance, extinction, and stellar type ($T_{\hbox{eff}}$, [Fe/H], $\log g$) information for 220 million stars.
Throughout this work, we denote the \citet{Zhang2023} catalog by ZGR23.

%
Compared to other stellar distance and extinction catalogs, the ZGR23 catalog features smaller uncertainties on the extinction estimates while still targeting a significant number of stars.
Approximately 87 million ZGR23 stars have an $A_V$ uncertainty below \SI{60}{mmag}.  
Thus, ZGR23 achieves similar extinction uncertainties compared to the subset of $\num[group-separator={,}]{39538}$ stars in the StarHorse catalog~\citep{Queiroz2023} that have both higher resolution spectra from the Apache point observatory galactic evolution experiment (APOGEE) and $grizy$ photometry from the panoramic survey telescope and rapid response system (Pan-STARRS), specifically Pan-STARRS1~\citep[PS1;][]{chambers2019} (typical $A_V$ extinction uncertainty of \SI{60}{mmag}).
While the ZGR23 catalog is limited to stars with \textit{Gaia} BP/RP measurements, the quality of the data makes the inference from the ZGR23 catalog competitive with models based on catalogs with larger numbers of stars --- 799 million stars in Bayestar19~\citep{Green2019}, 265 million in StarHorse DR2~\citep{Anders2019}, and 362 million in StarHorse EDR3~\citep{Anders2022}.
We further find the ZGR23 catalog to have fewer systematic shifts in the extinction and reliable extinction uncertainties based on an analysis in dust-free regions; further details are given in \Cref{appx:zgr23_in_dust_free_regions}.

For our reconstruction, we restricted our analysis to ZGR23 stars that have \texttt{quality\_flags$<$8}, as recommended by the authors.
We further sub-selected the stars based on their distance.
We required $\nicefrac{1}{(\omega-\sigma_\omega)}<\SI{1.8}{kpc}$ and $\nicefrac{1}{(\omega+\sigma_\omega)}>\SI{40}{pc}$ with $\omega$ the parallax of a star and $\sigma_\omega$ the parallax uncertainty to enforce that all stars are likely within our reconstructed volume.
In total, we selected \num[group-separator={,}]{53880655} stars.

\begin{figure}[!htbp]
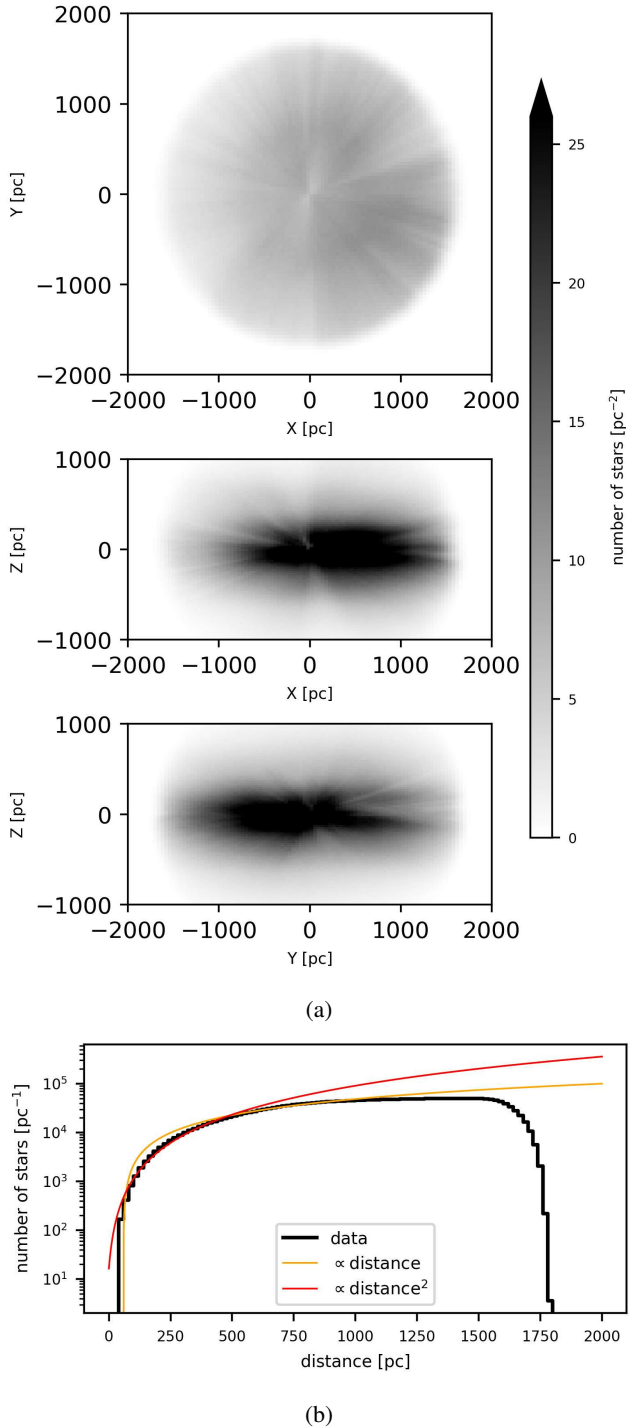

   \subcaptionbox{}{%
      \includegraphics[width=0.95\hsize,keepaspectratio]{{{res/density_of_stars}}}
   }
   \subcaptionbox{}{%
      \includegraphics[width=0.95\hsize,keepaspectratio]{{{res/selection_function}}}
   }
   \caption{%
      2D histograms of the density of stars in heliocentric Galactic Cartesian (X, Y, Z) projections, as well as the density of stars as a function of distance, for the subset of the ZGR23 catalog used in the reconstruction of our 3D dust map.
      Panel (a): Heliocentric Galactic Cartesian-projected histograms.
      Panel (b): Number of stars as a function of distance.
      This panel also shows a linear growth and a quadratic growth with distance for comparison.
      \label{fig:density_of_stars}
   }
\end{figure}

\begin{figure}[!htbp]
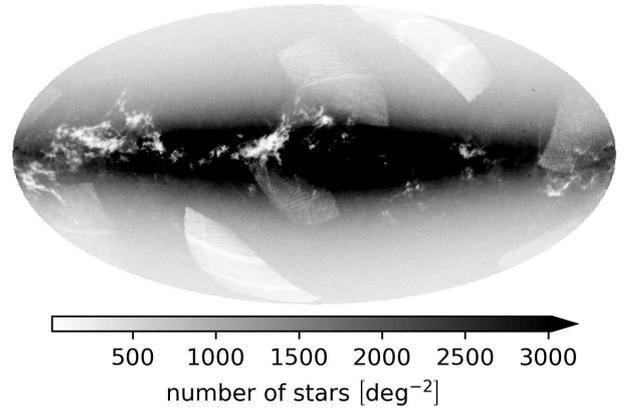

   \centering
   \includegraphics[width=0.95\hsize,keepaspectratio]{{{res/density_of_stars_mollview}}}
   \caption{%
      POS distribution of the subset of ZGR23 stars used in the reconstruction of our 3D dust map.
      \label{fig:density_of_stars_mollview}
   }
\end{figure}

The reliability of our reconstruction is predominantly limited by the quality and quantity of the data.
Both strongly depend on the POS position and distance.
\Cref{fig:density_of_stars} shows 2D histograms of stellar density in heliocentric Galactic Cartesian (X, Y, Z) projections, as well as the number of stars as a function of distance.
The densities of stars per distance bin first increases approximately quadratically with distance before falling off to a linear increase.
At approximately $\SI{1.5}{kpc}$ the number of stars per distance bin levels off due to our requirement that stars have a >1 sigma chance of being within \SI{1.8}{kpc} in distance.
\Cref{fig:density_of_stars_mollview} shows a POS histogram of the stars.
A clear imprint of the \textit{Gaia} BP/RP selection function is visible (cf. \citealt{CantatGaudin2022}).
A systematic under-sampling of stars behind dense dust clouds is also apparent.
We expect our reconstruction to be more trustworthy in regions of higher stellar density.
Due to the obscuring effect of dust, regions within and behind dense dust clouds should be treated with more caution.

\section{Priors}
\label{sec:priors}

Our quantity of interest is the 3D distribution of differential ZGR23 extinction $\rho$.\footnotemark{}
By definition, the differential extinction is positive.
Furthermore, we assumed it to be spatially smooth.
A priori we assumed the level of smoothness to be spatially stationary and isotropic.
\footnotetext{%
   The ZGR23 extinction is in arbitrary units but can be translated to an extinction at any given wavelength by using the extinction curve published at \url{https://doi.org/10.5281/zenodo.7692680}.
   Furthermore, dust extinction can be translated to a rough hydrogen volume density by assuming a constant extinction to hydrogen column density ratio \citep[see, e.g.,][]{Zucker2021}.
}

To reconstruct the 3D volume efficiently, we discretized it in spherical coordinates.
Specifically, we discretized our reconstructed volume into HEALPix spheres at logarithmically spaced distances.
We adopted an $N_\text{side}$ of $256$, which corresponds to $786,432$ POS bins.
This $N_\text{side}$ corresponds to an angular size of our voxels of $14'$.  
For the LOS direction, we adopted $772$ logarithmically spaced distance bins, of which $256$ are used for padding.
Our highest distance discretization is \SI{0.4}{pc} and our lowest distance discretization is \SI{7}{pc}.
In contrast to reconstructions with linearly spaced voxels in distance, we were able to probe much larger volumes while maintaining a high sampling at nearby distances.
The discretization provides a lower bound on the minimum separation between dust structures that we are able to resolve.
In practice the resolvable separation depends on the quantity and quality of the data and varies with the POS and LOS position.

We encoded both positivity and smoothness in our model by assuming the differential extinction to be log-normally distributed:
\begin{equation}
   \rho = \exp{s}
   ,\end{equation}
with normally distributed $s$, where $s$ is drawn from a GP with a homogeneous and isotropic correlation kernel, $k$.
From previous reconstructions of the differential extinction for the \textit{Gaia} DR2 G-band $A_G$ \citep{Leike2020}, we have constraints on the correlation kernel of the logarithm of the differential extinction in a volume around the Sun ($|X| < \SI{370}{pc},\, |Y| < \SI{370}{pc},\, |Z| < \SI{270}{pc}$).
As part of our prior model, we used the inferred $A_G$ extinction kernel from~\citet{Leike2020}.
To account for the conversion between the ZGR23 extinction and $A_G$ extinction, we added a global multiplicative factor to $s$ in our model.\footnotemark{}
Furthermore, we inferred an additive offset in the differential extinction.
We placed a log-normal prior on the multiplicative parameter and a normal prior on the additive one.
\footnotetext{%
   By doing so (and by using ZGR23) we implicitly assumed a spatially stationary reddening law for dust.
}

We enforced the correlation kernel $k$ using iterative charted refinement \citep[ICR;][]{Edenhofer2022}.
ICR enables us to enforce a kernel on arbitrarily spaced voxels by representing the modeled volume at multiple discretizations.
It starts from a very coarse view of our modeled volume.
On this coarsest scale, ICR models the GP with learned voxel excitations $\xi_e^{(0)}$ and an explicit full kernel covariance matrix.
A priori the parameters $\xi_e^{(0)}$ are standard normally distributed and coupled according to $k$ via ICR.
It then iteratively refines $n_\text{lvl}$ times its coarse view of the space with local, fine, a priori standard normally distributed corrections $\xi_e^{(1)},\dots,\xi_e^{(n_\text{lvl})}$ until reaching the desired discretization.
In each refinement, it uses $n_\mathrm{csz}$ neighbors from the previous refinement to refine one coarse pixel into $n_\mathrm{fsz}$ fine pixels (see \Cref{alg:icr}).

\begin{algorithm}
   \SetAlgoLined
   \DontPrintSemicolon
   \KwIn{$\xi_e^{(0)},\dots,\xi_e^{(n_\mathrm{lvl})}$}
   \Parameter{$\text{kernel}, n_\mathrm{csz}, n_\mathrm{fsz}$}
   \KwResult{$s^{(n_\mathrm{lvl})}$}

   \SetKwProg{Def}{def}{:}{}

   \Def{indices\_of\_neighbors(index, level, n\_neighbors)}{
      $\dots$\;
      \Return indices\;
   }
   \Def{cartesian\_positions(index, level, n\_neighbors)}{
      $\dots$\;
      \Return positions\;
   }
   \Def{refine(coarse, fine, $p_c$, $p_f$, kernel)}{
      $k_{cc}$ $\gets$ kernel($p_c$, $p_c$)\;
      $k_{ff}$ $\gets$ kernel($p_f$, $p_f$)\;
      $k_{cf}$ $\gets$ kernel($p_c$, $p_f$)\;
      $r$ $\gets$ $k_{cf}^T \cdot k_{cc}^{-1}$\;
      $d$ $\gets$ $k_{ff} - k_{cf}^{T} \cdot k_{cc}^{-1} \cdot k_{cf}$\;
      $\sqrt{d}$ $\gets$ cholesky($d$)\;
      \Return $r$ $\cdot$ coarse + $\sqrt{d}$ $\cdot$ fine\;
   }

   $s^{(0)}$ $\gets$ explicit\_gp(kernel, $\xi_e^{(0)}$)\;
   \For{$l \gets 1$ \KwTo $n_\mathrm{lvl}$}{
   \For{$i$ \KwInIterable $\mathrm{ndindex}(\mathrm{shape}(s^{(l-1)}))$}{
   $c$ $\gets$ indices\_of\_neighbors($i$, $l-1$, $n_\mathrm{csz}$)\;
   $f$ $\gets$ indices\_of\_neighbors($i$, $l$, $n_\mathrm{fsz}$)\;
   $p_c$ $\gets$ cartesian\_positions($i$, $l-1$, $n_\mathrm{csz}$)\;
   $p_f$ $\gets$ cartesian\_positions($i$, $l$, $n_\mathrm{fsz}$)\;
   $s^i[f]$ $\gets$ refine($s^{(i-1)}[c]$, $\xi_e^{(i)}[f]$, $p_c$, $p_f$, kernel)\;
   }
   }

   \caption{%
      Pseudocode for ICR creating a GP $s$ from uncorrelated excitations $\left\{ \xi_e^{(0)},\dots,\xi_e^{(n_\mathrm{lvl})} \right\}$.
      Each coarse pixel at location $j$ is iteratively refined to $n_\mathrm{fsz}$ fine pixels using $n_\mathrm{csz}$ coarse pixel neighbors.
      The correlation kernel is denoted by $k$.
      Square brackets after variables and the two functions \texttt{ndindex} and \texttt{shape} denote NumPy-like \citep{Harris2020} indexing routines.
      The call \texttt{explicit\_gp} refers to an unspecified Gaussian Process model explicitly representing the covariance of $k$ for the pixel positions modeled by $\xi^{(0)}_e$.
      \label{alg:icr}
   }
\end{algorithm}

Iterative charted refinement uses local corrections at varying discretizations and within a refinement assumes the previous iteration to have modeled the GP without error.
Both lead to slight errors in representing the kernel.
For our use case, we encountered errors in representing the kernel of a few percent.
We accepted these errors as a trade-off that enables the reconstruction to probe larger volumes.
We refer to \citet{Edenhofer2022} for a detailed discussion of the kernel approximation errors.

Overall, our model for the prior reads
\begin{equation}
   \rho = \exp{\left[\mathrm{scl}(\xi_\mathrm{scl}) \cdot s\left(\xi_e^{(0)},\dots,\xi_e^{(n_\mathrm{lvl})}\right) + \mathrm{off}(\xi_\mathrm{off})\right]},
\end{equation}
where we denote the learned multiplicative scaling of $s$ by $\mathrm{scl}$, the learned additive offset by $\mathrm{off}$, and re-expressed both in terms of a priori standard normally distributed parameters $\xi_\mathrm{scl}$ and $\xi_\mathrm{off}$, respectively.
The act of expressing $\mathrm{scl}$, $\mathrm{off}$ and $s$ via parameters with an a priori simpler distribution, here a standard normal distribution, is called re-parameterization.
A detailed discussion on this subject is given in \citet{Rezende2015}.

\section{Likelihood}
\label{sec:likelihood}

To construct the likelihood we first needed to define how the differential extinction $\rho$ --- our quantity of interest --- connects to the measured data $\mathcal{D}$.
Our data comprise POS position, extinction $\mathcal{D}_A$, and parallax $\mathcal{D}_\omega$ data.
The POS position is in essence without error.
The extinction data $\mathcal{D}_A=\{A,\sigma_A\}$ are in the form of integrated LOS extinctions to stars $A$ and associated uncertainties $\sigma_A$
The parallax data $\mathcal{D}_\omega=\{\omega, \sigma_\omega\}$ similarly are in the form of parallax estimates $\omega$ and uncertainties $\sigma_\omega$.

Our model focuses on the measured extinction, $A$, and does not predict parallaxes to stars.
Instead, we conditioned our model on the parallax data, $\mathcal{D}_\omega,$ and split the likelihood into the probability of the measured extinction given the true extinction, $a$, and the probability of the true extinction given uncertain parallax information:
\begin{align}
   P(A\,|\,\rho,\mathcal{D}_\omega) & = \int\mathrm{d}{a}\ P(A,a\,|\,\rho,\mathcal{D}_\omega)
   \\ &= \int\mathrm{d}{a}\ P(A\,|\,a) \cdot P(a\,|\,\rho,\mathcal{D}_\omega)\ . \label{eq:top_level_likelihood}
\end{align}
The first term of the integrand is constrained by the quality of the extinction measurements and the second by the quality of the parallax measurements.

\subsection{Response}
\label{sec:likelihood:response}

The second term in \Cref{eq:top_level_likelihood}, $P(a\,|\,\rho,\mathcal{D}_\omega)$, can be expressed as the joint probability of extinction and true distance, $d$, marginalized over the true distance:
\begin{align}
   P(a\,|\,\rho,\mathcal{D}_\omega)
    & = \int\mathrm{d}d\ P(a,d\,|\,\rho,\mathcal{D}_\omega)
   \\ &= \int\mathrm{d}d\ P(a\,|\,\rho,\mathcal{D}_\omega,d) \cdot P(d\,|\,\rho,\mathcal{D}_\omega)
   \ .
\end{align}
We neglected data selection effects (i.e., $a$'s dependence on $\mathcal{D}_\omega$ given $d$ and $d$'s dependence on $\rho$ given $\mathcal{D}_\omega$) and used the fact that the true extinction, $a$, at known distance $d$ is simply the LOS integral of $\rho$ along the LOS to the star from zero to $d$:
\begin{align}
   P(a|\rho,\mathcal{D}_\omega)
    & = \int\mathrm{d}d\ P(a|\rho,d) \cdot P(d|\mathcal{D}_\omega)
   \\ &= \int\mathrm{d}d\ \delta\left( a - \underbrace{\int_0^{d}\mathrm{d}\tilde{d}\ \rho[\mathrm{POS}](\tilde{d})}_{\coloneqq R^{d}(\rho)} \right) \cdot P(d|\mathcal{D}_\omega),
\end{align}
with $\rho[\mathrm{POS}]$ the slice of $\rho$ at the POS positions of the stars, $\delta$ the Dirac delta distribution defined by $\int_{-\infty}^{\infty}\mathrm{d}x\ f(x)\delta(x)=f(0)$ for any continuous $f$ with compact support, and $R$ the response that maps from $\rho$ to the domain of the measured extinction.

We approximated $P(a|\rho,\mathcal{D}_\omega)$ with a normal distribution,
\begin{equation}
   P(a|\rho,\mathcal{D}_\omega) \approx \mathcal{G}\left(a|\bar{a},\sigma_a^2\right)
   ,\end{equation}
with mean $\bar{a}$ and standard deviation $\sigma_a$ to obtain a tractable expression for \Cref{eq:top_level_likelihood}.
The mean extinction, $\bar{a}$, is
\begin{align}
   \bar{a} & \coloneqq {\langle a \rangle}_{P(a|\rho,\mathcal{D}_\omega)}
   \\ &= \int\mathrm{d}a\,a\int\mathrm{d}d\ \delta\left( a - R^{d}(\rho) \right) \cdot P(d|\mathcal{D}_\omega)
   \\ &= \int\int\mathrm{d}a\,\mathrm{d}d\ a \cdot \delta\left( a - R^{d}(\rho) \right) \cdot P(d|\mathcal{D}_\omega)
   \\ &= \int\mathrm{d}d\ R^{d}(\rho) \cdot P(d|\mathcal{D}_\omega)
   \\ &= {\left\langle R^{d}(\rho) \right\rangle}_{P(d|\mathcal{D}_\omega)}
   \ .
\end{align}
Assuming the parallax $\nicefrac{1}{d}$ is normally distributed (i.e., $P(d|\mathcal{D}_\omega)=\mathcal{G}\left(\nicefrac{1}{d} \vert \omega, \sigma_\omega^2\right)$ with mean $\omega$ and standard deviation $\sigma_\omega$), then
\begin{align}
   \begin{split}
      {\langle a \rangle}_{P(a|\rho,\mathcal{D}_\omega)} &= {\left\langle R^{d}(\rho) \right\rangle}_{\mathcal{G}(\nicefrac{1}{d} \vert \omega, \sigma_\omega^2)}
      \\ &= \int_0^{\infty}\mathrm{d}\tilde{d}\ \rho[\mathrm{POS}](\tilde{d}) \cdot \mathrm{sf}_\mathcal{G}\left(\nicefrac{1}{\tilde{d}} \vert \omega, \sigma_\omega^2\right),
   \end{split}
\end{align}
with $\mathrm{sf}_\mathcal{G}\left(\nicefrac{1}{d} \vert \omega, \sigma_\omega^2\right) \coloneqq 1 - \int_{-\infty}^{\nicefrac{1}{d}}\mathrm{d}\omega'\ \mathcal{G}\left(\omega' \vert \omega, \sigma_\omega^2\right)$ the survival function of the normal distributed parallax.

The standard deviation $\sigma_{a}$ can be understood as an additional error contribution for marginalizing over the distance.
The error depends on the distance uncertainty and the dust along the full LOS:
\begin{equation}
   \sigma_{a}^2 \coloneqq {\left\langle {\left(R^{d}(\rho)\right)}^2 \right\rangle}_{\mathcal{G}\left(\nicefrac{1}{d} \vert \omega, \sigma_\omega^2\right)} - {\langle R^{d}(\rho) \rangle}^2_{\mathcal{G}\left(\nicefrac{1}{d} \vert \omega, \sigma_\omega^2\right)}
   \ .
\end{equation}
Evaluating both $\bar{a}$ and $\sigma_a^2$ is comparatively cheap in a spherical coordinate system since for a discretized sphere $R^{d}(\rho)$ is simply the cumulative sum of $\rho$ along the distance axis weighted by the radial extent of each voxel.

\subsection{Likelihood and joint probability density}

We assumed the measured extinction to be normally distributed around the true extinction $a$.
We took the inferred extinction, $A$, from the catalog to be the mean of the normal distribution.
The accompanying uncertainty $\sigma_A$ in the catalog was assumed to be the standard deviation of $P(A\,|\,a)$.

Some stars will have underestimated uncertainties due to either mismodeled intrinsic stellar properties in the inference or bad photometric measurements that were not flagged.
We wanted our model to be able to detect and deselect stars that are in strong disagreement with the rest of the reconstruction.
We achieved this by inferring an additional multiplicative factor per star, $n_\sigma$, which scales $\sigma_A$.
A priori, we assumed $n_\sigma$ to be drawn from a heavy-tailed distribution.
Specially, we assumed $n_\sigma$ to follow an inverse gamma distributed.
We again express $n_\sigma$ in terms of standard normally distributed parameters $n_\sigma(\xi_\sigma)$ in the inference.

To summarize, our approximate likelihood first introduced in \Cref{eq:top_level_likelihood}, reads
\begin{align}
   \begin{split}
      &P(A\,|\,\rho,n_\sigma,\mathcal{D}_\omega)
      \\ &\approx \int\mathrm{d}{a}\ \mathcal{G}\left(A\,\left|\,a,{\left(n_\sigma \cdot \sigma_A\right)}^2\right.\right) \cdot \mathcal{G}\left(a\,\left|\,\bar{a}(\rho),\sigma_a^2(\rho)\right.\right)
   \end{split}
   \\ &= \mathcal{G}\left( A \,\left|\,\bar{a}(\rho),\left[n_\sigma \cdot \sigma_A\right]^2+\sigma_a^2(\rho) \right.\right). \label{eq:total_likelihood}
\end{align}
The uncertainty in the extinction $\sigma_{A}$ is scaled by $n_\sigma$ to deselect outliers and increased by $\sigma_{a}^2$ due to marginalizing over the distance uncertainty.

The joint probability density function of data and parameters reads
\begin{align}
   \begin{split} \label{eq:joint_model}
      P&(A,\rho(\xi),n_\sigma(\xi) \vert \mathcal{D}_\omega)
      \\ &=\ \mathcal{G}\left( A \,\left|\,\bar{a}(\rho(\xi)),\left[n_\sigma(\xi) \cdot \sigma_A\right]^2+\sigma_a^2(\rho(\xi)) \right.\right)
      \cdot \mathcal{G}\left( \xi \,\left|\,0, 1\right.\right),
   \end{split}
\end{align}
with $\xi$ the vector of all parameters of the model $\left\{\xi_e^{(0)},\dots,\xi_e^{(n_\mathrm{lvl})},\xi_\mathrm{scl},\xi_\mathrm{off},\xi_\sigma\right\}$.
The complexity of the prior distributions has been fully absorbed into the transformations $s(\xi)$, $\mathrm{scl}(\xi)$, $\mathrm{off}(\xi)$, and $n_\sigma(\xi)$ from the a priori standard normally distributed parameters $\xi$.

\begin{table*}[!htbp]
   \caption{%
      Parameters of the prior distributions.
      The parameters $s$, $\mathrm{scl}$, and $\mathrm{off}$ fully determine $\rho$.
      They are jointly chosen to a priori yield the kernel reconstructed in \citet{Leike2020}.
   }
   \label{tab:priors}
   \centering
   \begin{tabular}{c c c c c}
      \hline\hline
      Name           & Distribution  & Mean                                                                                                                       & Standard Deviation            & Degrees of Freedom                             \\
      \hline
      $s$            & Normal        & $0.0$                                                                                                                      & Kernel from \citet{Leike2020} & $\num[group-separator={,}]{786432} \times 772$ \\
      $\mathrm{scl}$ & Log-Normal    & $1.0$                                                                                                                      & $0.5$                         & 1                                              \\
      $\mathrm{off}$ & Normal        & \begin{tabular}{@{}c@{}} $-6.91\left(\approx\ln10^{-3}\right)$ \\ prior median extinction \\ from \citet{Leike2020} \end{tabular} & $1.0$                         & 1                                              \\
      \hline
                     &               & Shape Parameter                                                                                                            & Scale Parameter               &                                                \\
      $n_\sigma$     & Inverse Gamma & $3.0$                                                                                                                      & $4.0$                         & \# Stars = \num[group-separator={,}]{53880655} \\
      \hline
   \end{tabular}
\end{table*}

Our priors in terms of nonstandard-normal parameters are summarized in \Cref{tab:priors}.
The priors for $s$, $\mathrm{scl}$, and $\mathrm{off}$ are chosen to a priori yield the kernel reconstructed in \citet{Leike2020}.
In contrast to \citet{Leike2020}, we do not learn a full non-parametric kernel.
However, we do infer $\mathrm{scl}$ and $\mathrm{off}$, the scale, and zero-mode of the kernel.
The prior for $n_\sigma$ was chosen such that the inverse gamma distribution has mode $1$ and standard deviation $2$.

\section{Posterior inference}
\label{sec:posterior_inference}

In the previous section we took special care to express our model not only in terms of physical parameters, like the differential extinction density $\rho$, but also in terms of simpler parameters $\xi$.
The act of expressing the parameters of the model $\mathrm{scl}$, $\mathrm{off}$, $s$, and $n_\sigma$ in terms of a priori standard normal distributed variables $\xi$ is called standardization, a special from of re-parameterization \citep[see][]{Rezende2015}.
Effectively, we are shifting complexity from the prior to the likelihood.
However, both the nonstandardized and the standardized formulation of the joint model are equivalent.
Standardizing models can lead to better conditioned inference problems as the parameters all vary on the same scales --- if the prior is not in conflict with the likelihood.
We used an inference scheme that relies on the standardized formulation.

We wanted to infer the posterior for our standardized model from \Cref{eq:joint_model}.
Directly probing the posterior via sampling methods such as Hamiltonian Monte Carlo \citet{Hoffman2011} is computationally infeasible.
Instead, we used variational inference to approximate the true posterior.
Specifically, we used metric Gaussian variational inference \citep[MGVI;][]{Knollmueller2019}.
We summarize the main idea behind MGVI in \Cref{appx:mgvi}.
We did not approximate the posterior of the noise inference parameter $n_\sigma(\xi_\sigma)$ via variational inference and instead used only the maximum of the posterior for $\xi_\sigma$.

To speed up the inference, we started the reconstruction at a lower resolution ($196,608$ POS bins at $N_\text{side}=128$ and $388$ LOS distance bins) and restricted the inference to a subset of stars with a $\geq 2$ sigma chance of being within \SI{600}{pc} and a $\geq 2$ sigma chance of being farther than \SI{40}{pc}.
We successively increased the distance range of the map up to which stars are incorporated in steps of \SI{300}{pc} from \SI{600}{pc} to \SI{1.8}{kpc}.
Every time we increased the distance range, we reset the parameters for $n_\sigma$.
Then, after all data were incorporated, we increased the angular and distance resolution of the reconstruction to the final resolution.

Our data selection deselects stars close to the maximum distance probed (c.f. \Cref{fig:density_of_stars}).
This effect leads to the outer regions of the map being informed by relatively few stars compared to the inner regions.
We observe that these regions are prone to producing spurious features.
For our final data products, we removed the outermost \SI{550}{pc} from the data-constrained volume as we observed artifacts aligned with our data incrementation strategy within these regions.
We believe \SI{550}{pc} to be a conservative cut but we advise caution when finding structures perfectly aligned with a sphere around the Sun at \SI{600}{pc}, \SI{900}{pc}, or \SI{1200}{pc}.

ZGR23 assumes all extinctions to be strictly positive.
We neglected this constraint by assuming Gaussian errors, which led to an artificial spike in extinction in the first few voxels in each direction.
As we know those regions to be effectively free of dust from previous reconstructions c.f.~\citet{Leike2020}, we removed the innermost \SI{69}{pc} (see \Cref{appx:extinction_within_the_innermost_x_pc}).
We release an additional HEALPix map of integrated extinction out to \SI{69}{pc} from the Sun and suggest using it to correct integrated LOS predictions for the removed extinction.

Our inference heavily utilizes derivatives of various components of our model.
Derivatives are used for the minimization as well as for the variational approximation of the posterior.
Previous models such as those described in \citet{Leike2019} and \citet{Leike2020} relied on the Numerical Information Field Theory (NIFTy) package \citep{Selig2013,Steiniger2017,Arras2019} and were limited to running on CPUs.

We employed a new framework called NIFTy.re \citep{Edenhofer2023NIFTyRE} for deploying NIFTy models to GPUs.
NIFTy.re is part of the NIFTy Python package and internally uses JAX~\citep{Jax2018} to run models on the GPU.
We were able to speed up the evaluation of the value and gradient of \Cref{eq:joint_model} by two orders of magnitude by transitioning from CPUs to GPUs.
Our reconstruction ran on a single NVIDIA A100 GPU with 80 GB of memory for about four weeks.

\section{Caveats}
\label{sec:caveats}

We believe statistical uncertainties are the dominant source of uncertainty for our reconstruction.
However, it is important to also consider sources of systematic uncertainties.
Depending on the application, the systematic uncertainties may be more important than the statistical uncertainties.
The data that informed the reconstruction, the model with which we inferred it, and the inference procedure all contribute to the model systematic uncertainties.

Naturally, the data themselves are a source of systematic uncertainties \citep[spatially stationary reddening law, mismodeling of binaries, etc.; see][]{Zhang2023} and additionally is known to be incomplete, c.f.~\Cref{fig:density_of_stars_mollview}.
A lower stellar density, for example in heavily obscured regions, limits the map's resolution and results in volumes of the map behind dense dust clouds being poorly constrained.
Thus, we believe our dust reconstruction to be an underestimation of the true extinction toward dense dust clouds.
\citet{Zucker2021} also note this effect when comparing the \citet{Leike2020} map with 2D integrated extinction maps based on infrared photometry, finding that the \citet{Leike2020} is not sensitive to regions with $A_V \gtrsim 2$ mag.

We advise visualizing the stellar density in the region of interest to assess the magnitude of the systematic uncertainties due to data incompleteness.
We release all stars used in the reconstruction as an additional data product.
This data product can be used to visualize the stellar density.
In regions with a significant underdensity of stars, we expect the systematic lack of stars to dominate the statistical uncertainties in the reconstruction.
The reconstruction produced plausible infills in those regions based on adjacent stars.
However, the resulting uncertainties do not capture the cause for the systematic lack of stars as the model implicitly assumes that underdensities of stars are not systematic but purely random.

Our model includes a number of approximations.
First, we assumed a GP prior on the logarithmic dust extinction using the kernel from \citet{Leike2020} and additionally only applied it approximately via ICR.
Second, we assumed $\mathcal{D}_\omega$ to be independent of $\mathcal{D}_A$.
Third, we assumed the parallax error to be Gaussian, and fourth, we assumed the extinction error to be Gaussian.

For extremely low extinctions, the assumption of $A$ being Gaussian is poor due to the positivity prior in the ZGR23 catalog.
We corrected for this bias toward higher estimated extinction in regions with assumed extremely low true extinctions post-hoc by cutting away the innermost \SI{69}{pc} as described in \Cref{sec:posterior_inference}.
We publish an auxiliary map of integrated extinction out to \SI{69}{pc} from the Sun to correct integrated LOS predictions for the removed extinction.
We suggest adding the removed local extinction back to the map when comparing integrated extinctions.
By default, the removed local extinction is added back to the map when querying integrated extinctions via \texttt{dustmaps}.

We further release a catalog of the predicted extinctions of our model to all stars that we used for the reconstruction.
In \Cref{appx:extinction_catalog}, we perform a non-exhaustive consistency test comparing our predictions to the ZGR23 ones.
We find that both predictions for the extinction to stars disagree below \SI{50}{mmag} and above \SI{4}{mag}; 34\% more stars than expected have larger (respectively smaller) extinction predictions compared to ZGR23.
We expect the range from \SI{50}{mmag} to \SI{4}{mag} to correspond to the range within which our map is reliable.
More details are provided in \Cref{appx:extinction_catalog}.

Furthermore, our posterior inference is an approximation.
We assume our approximation of the true posterior accurately captures the intrinsic model uncertainties \citep[c.f.][]{Arras2022,Leike2019,Leike2020,Mertsch2023,Roth2023DirectionDependentCalibration,Hutschenreuter2023,Tsouros2023,Roth2023FastCadenceHighContrastImaging,Hutschenreuter2022}.
However, we worried about structures getting burned in when we increase the maximum distance probed during the inference from \SIrange{600}{1800}{pc} in steps of \SI{300}{pc} as described in \Cref{sec:posterior_inference}.
We checked the final reconstruction for this effect by comparing it against a larger reconstruction that does not sub-select the stars based on their distance during the inference but uses only a small subsample of ZGR23 stars with more stringent quality flags.
The larger reconstruction, which extends out to \SI{2}{kpc} in distance, is released as an additional data product.
We find no significant differences between both runs.
Details on the larger reconstruction are provided in \Cref{appx:2kpc_reconstruction}.

\section{Results}
\label{sec:results}

We reconstructed 12 samples (6 antithetically drawn samples) of the 3D dust extinction distribution, each of which encompasses \num[group-separator={,}]{607125504} differential extinction voxels.
The voxels are arranged on $772$ HEALPix spheres with $N_\text{side}=256$ spaced at logarithmically increasing distances.
After removing the innermost \SI{<69}{pc} and outermost \SI{>1250}{pc} HEALPix spheres, we are left with $516$ HEALPix spheres.
The samples, the posterior mean, and the posterior standard deviation for the reconstruction are publicly available online\footnotemark{}.
We strongly advise using the samples for any quantitative analysis.
For convenience, we also provide the posterior mean and standard deviation of the reconstruction interpolated to heliocentric Galactic Cartesian Coordinates (X, Y, Z) and Galactic spherical Coordinates ($l$, $b$, $d$) as well as the scripts for the interpolation.
Furthermore, the map can be queried via the \texttt{dustmaps} Python package~\citep{Green2018Dustmaps}.
Further details on using the reconstruction are given in \Cref{appx:using_the_reconstruction}.
\footnotetext{
	\url{https://doi.org/10.5281/zenodo.8187942}
}

The distance discretization in our reconstruction is highest for close-by voxels and decreases further out.
Our highest distance discretization is \SI{0.4}{pc} and our lowest distance discretization is \SI{7}{pc} while our angular discretization is $14'$ and is independent of the distance.
The stated discretizations specify the lower bound on our resolution.
The minimum separation that we are able to resolve depends on the position and is encoded in the posterior samples.
For small regions, we suggest additionally analyzing the stellar density (see \Cref{sec:caveats}) to assess the strength of the systematic uncertainties due to the density of stars.

The reconstruction is in terms of the unitless ZGR23 extinction as defined in~\citet{Zhang2023}.
For visualization purposes, we translated the ZGR23 extinction to Johnson's V-band $\lambda=\SI{540.0}{\nano\meter}$, that is to say, \ $A_V \coloneqq A(V=\SI{540.0}{\nano\meter})$.
To perform the conversion, we adopted the extinction curve published in ZGR23 and multiplied the unitless ZGR23 extinction by a factor of $2.8$.
We refer readers to the full extinction curve\footnotemark{} from \citet{Zhang2023} for the coefficients needed to translate the extinction to other bands.
\footnotetext{
	\url{https://doi.org/10.5281/zenodo.7692680}
}

\begin{figure*}[!htbp]
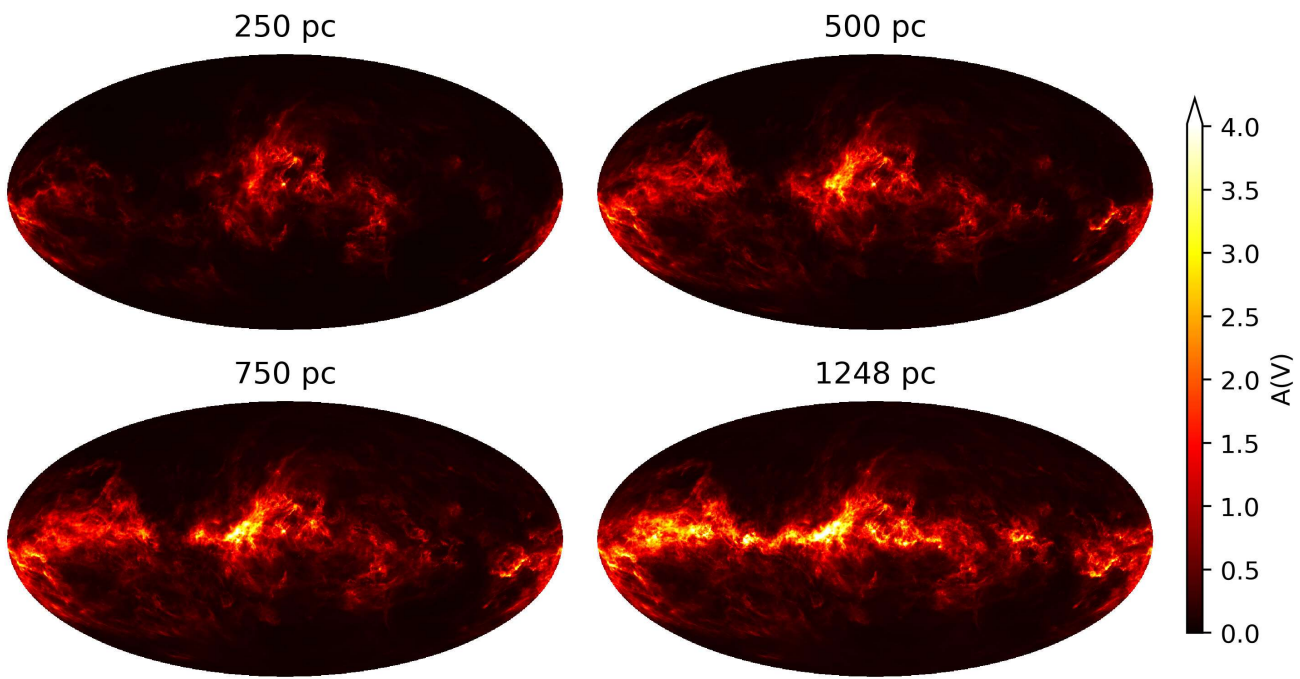

   \centering
   \includegraphics[width=0.95\hsize,keepaspectratio]{{{res/integrated_mollview}}}
   \caption{%
      Mollweide projection of the POS integrated $A_V$ extinction out to \SI{250}{pc}, \SI{500}{pc}, \SI{750}{pc}, and up to the maximum distance of our map.
      The colorbar saturates at the $99.9\%$ quantile.
      \label{fig:integrated_mollview}
   }
\end{figure*}

\begin{figure*}[!htbp]
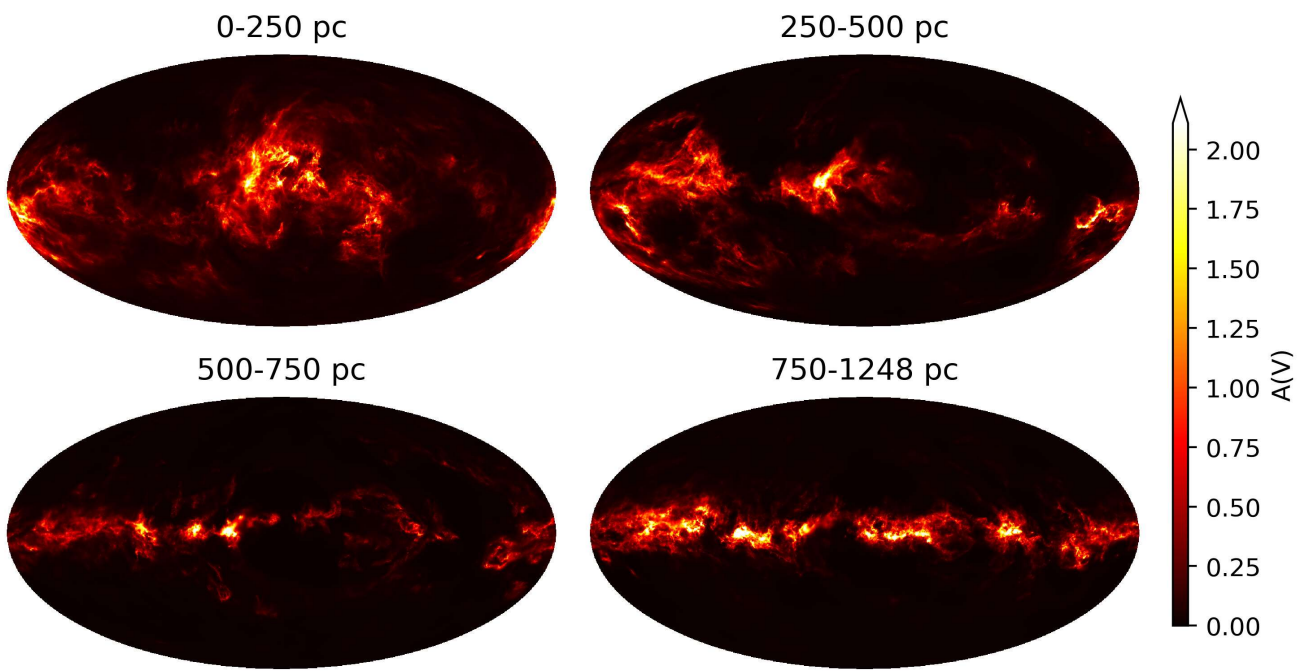

   \centering
   \includegraphics[width=0.95\hsize,keepaspectratio]{{{res/diff_mollview}}}
   \caption{%
      Same as \Cref{fig:integrated_mollview} but showing the difference between the integrated extinctions in between distance slices projected on the POS.
      The colorbar saturates at the $99.9\%$ quantile.
      \label{fig:diff_mollview}
   }
\end{figure*}

\Cref{fig:integrated_mollview} depicts the POS projection of the posterior mean reconstruction integrated out to \SI{250}{pc}, \SI{500}{pc}, \SI{750}{pc}, and up to the end of our sphere.
The $A_V$ values are in units of magnitudes.
We see that higher-latitude features like the Aquila Rift are comparatively close-by while structures in the Galactic plane appear only gradually.
\Cref{fig:diff_mollview} shows the difference between the integrated POS projections.
We recover well-known features of integrated dust but are now able to de-project them.

\begin{figure*}[!htbp]
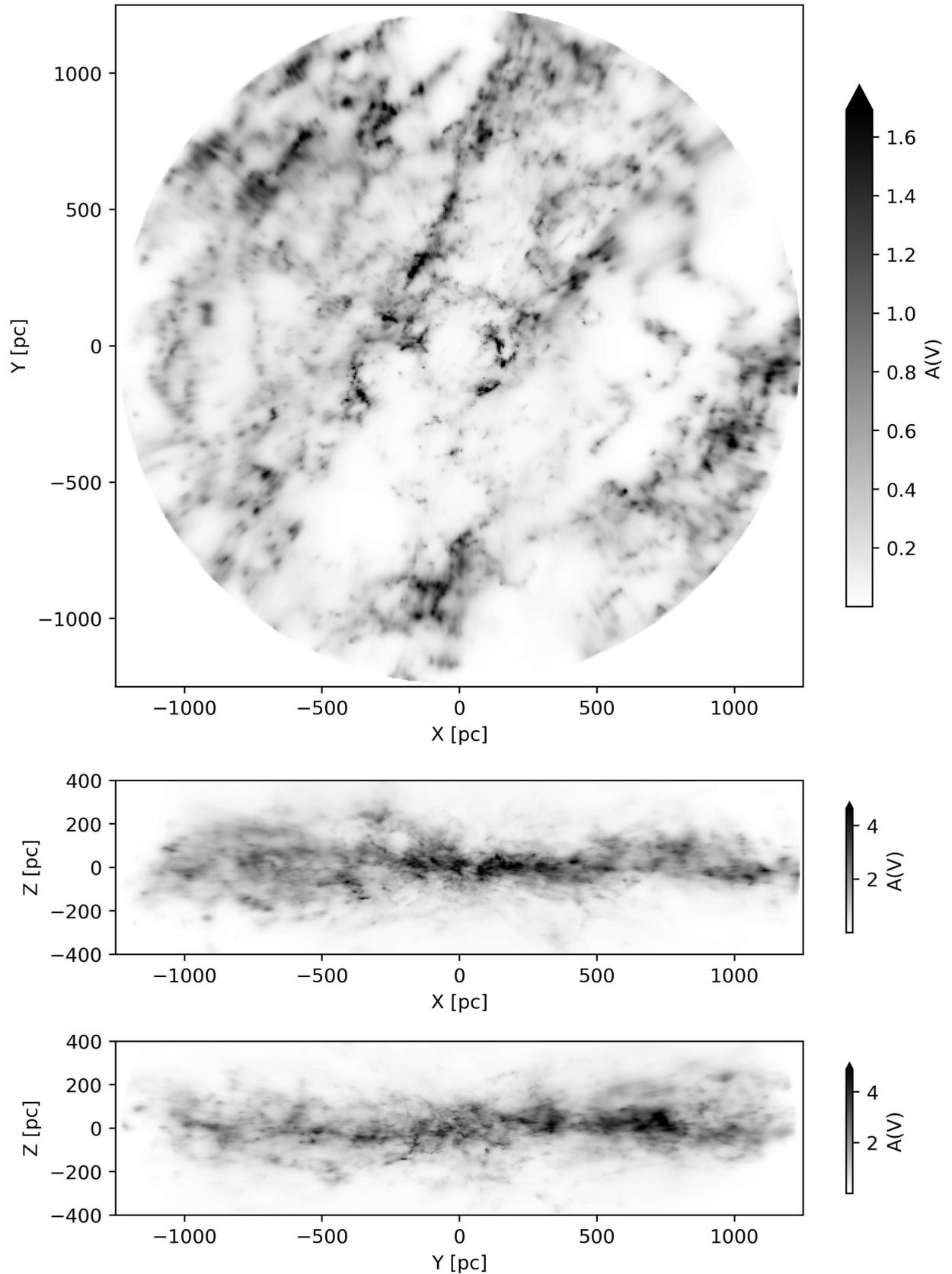

   \centering
   \includegraphics[width=0.9\hsize,keepaspectratio]{{{res/galactic}}}
   \caption{%
      Heliocentric Galactic Cartesian (X, Y, Z) projections of the posterior mean of our 3D dust map in a box with dimensions $\SI{2.5}{kpc} \times \SI{2.5}{kpc} \times \SI{0.8}{kpc}$ centered on the Sun.
      The colorbar is linear and saturates at the $99.9\%$ quantile.
      A GIF of the posterior samples is shown at \url{https://faun.rc.fas.harvard.edu/gedenhofer/perm/E+23/21b9_final.gif}.
      A low-resolution 3D interactive version of this figure is available at \url{https://faun.rc.fas.harvard.edu/czucker/Paper_Figures/3D_Dust_Edenhofer2023.html}.
      \label{fig:galactic}
   }
\end{figure*}

\begin{figure*}[!htbp]
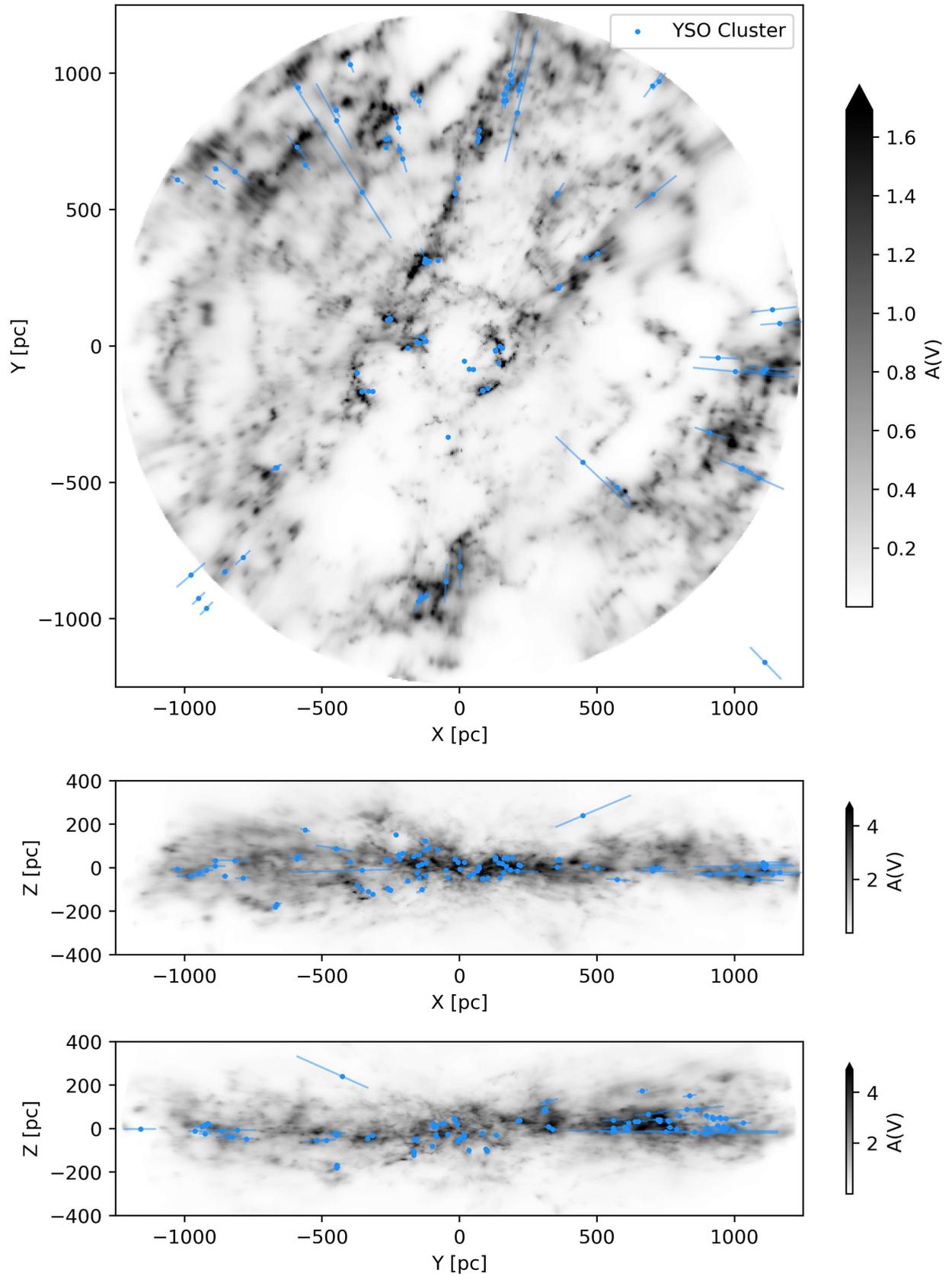

   \centering
   \includegraphics[width=0.9\hsize,keepaspectratio]{{{res/galactic_with_ysos}}}
   \caption{%
      Same as \Cref{fig:galactic} but with a catalog of clusters of YSOs~\citep{Kuhn2023YSO} based on \citet{Kuhn2021}, \citet{Winston2020}, and \citet{Marton2022} shown as blue dots on top of the reconstruction; their distance uncertainties are shown as extended lines.
      \label{fig:galactic_ysos}
   }
\end{figure*}

\begin{figure*}[!htbp]
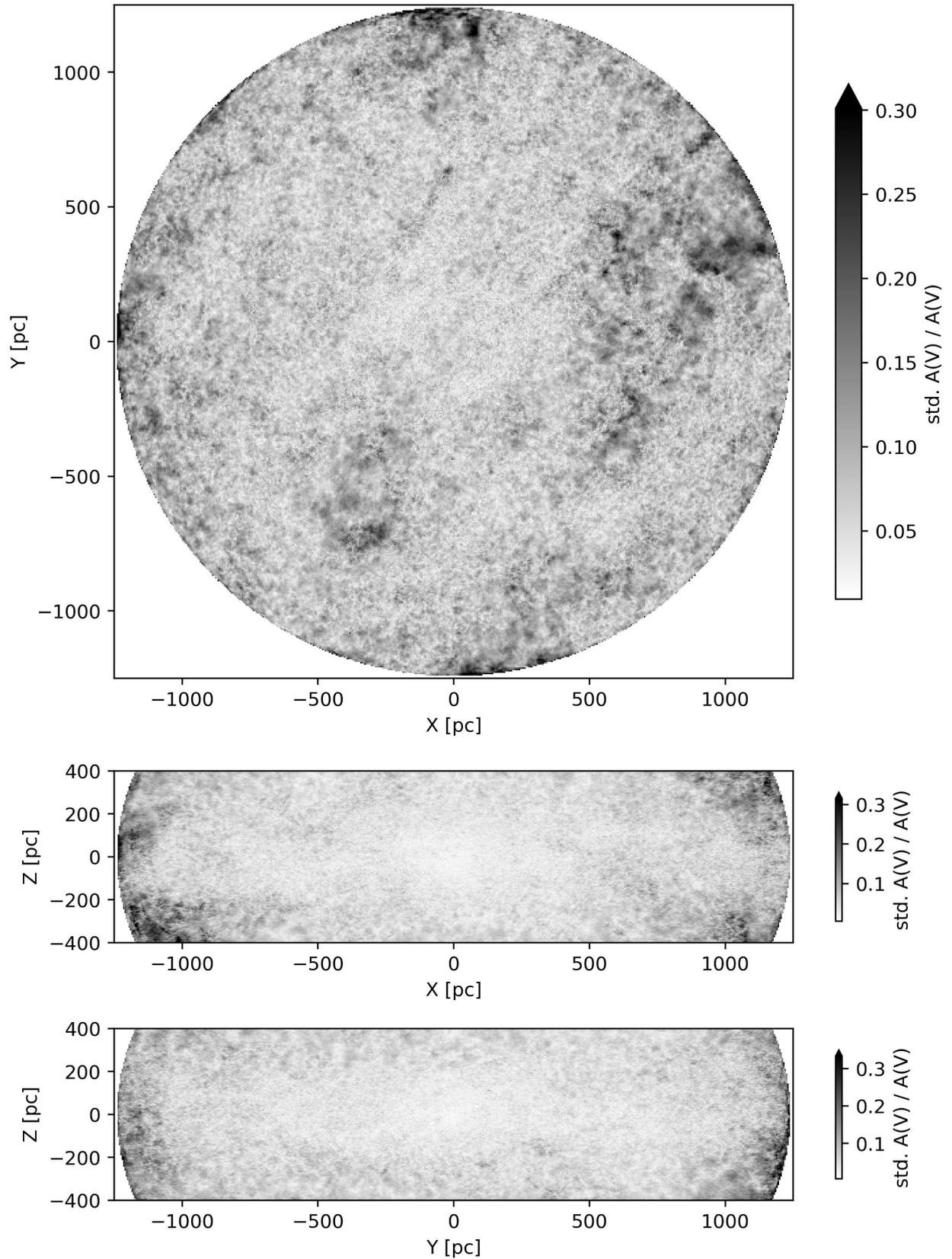

   \centering
   \includegraphics[width=0.9\hsize,keepaspectratio]{{{res/galactic_rel_std}}}
   \caption{%
      Heliocentric Galactic Cartesian (X, Y, Z) projections of the relative uncertainty of the reconstructed dust extinction integrated within a box of $\SI{2.5}{kpc} \times \SI{2.5}{kpc} \times \SI{0.8}{kpc}$ centered on the Sun.
      The colorbar is linear and saturates at the $99.9\%$ quantile.
      \label{fig:galactic_rel_std}
   }
\end{figure*}

\Cref{fig:galactic} show a bird's eye (X, Y), side-on (X, Z), and (Y, Z) projection of the posterior mean of our reconstruction in heliocentric Galactic Cartesian coordinates.
The image depicts the innermost $\SI{2.5}{kpc} \times \SI{2.5}{kpc} \times \SI{0.8}{kpc}$ around the Sun in $A_V$ extinction integrated over $z$ from \SIrange{-400}{400}{pc}, $y$ in \SIrange{-1.25}{1.25}{kpc}, and $x$ in \SIrange{-1.25}{1.25}{kpc}, respectively.
In \Cref{fig:galactic_ysos} we overlay a catalog of clusters of young stellar objects \citep[YSOs;][]{Kuhn2023YSO} based on \citet{Kuhn2021}, \citet{Winston2020}, and \citet{Marton2022}, which are shown as blue dots.
The positions of the YSO clusters visually agree with the positions of dust clouds within the YSO clusters' reported distance uncertainties.

The posterior standard deviation divided by the posterior mean of the reconstruction is shown in \Cref{fig:galactic_rel_std}.
The map features a faint speckle pattern.
This is likely due to the low number of samples relative to the number of degrees of freedom.
The standard deviation is on the order of $10\%$ of the posterior mean and slightly increases with distance.
Toward the galactic center behind the dust in the immediate vicinity of roughly \SI{300}{pc}, the relative uncertainty is noticeably higher.

The reconstruction has a high dynamic range and reveals faint dust lanes in the reconstructed volume.
Small approximately spherical cavities are evident throughout the map.
The dust clouds in the reconstruction are compact and only weakly elongated radially.
Prominent large-scale features, such as the Radcliffe Wave \citep{Alves2020} and the Split \citep{Lallement2019}, have been resolved at an unprecedented level of detail, previously only accessible for the most nearby dust clouds.

\subsection{Comparison to existing 3D dust maps}

In this section, we compare our map to other 3D dust maps in the literature.
We denote the dust map described in \citet{Leike2020} by LGE20, \citet{Vergely2022} by VLC22, \citet{Green2019} by Bayestar19, and \citet{Leike2022} by L+22 in this section.
For the purposes of comparison, we show the posterior mean.
We release the statistical uncertainties as additional data products, and we strongly advise taking into account the released statistical uncertainties for any quantitative analysis.
However, the differences between the various 3D dust reconstructions discussed here are systematic differences and are not captured by the reconstructed statistical uncertainties.

\begin{figure*}[!htbp]
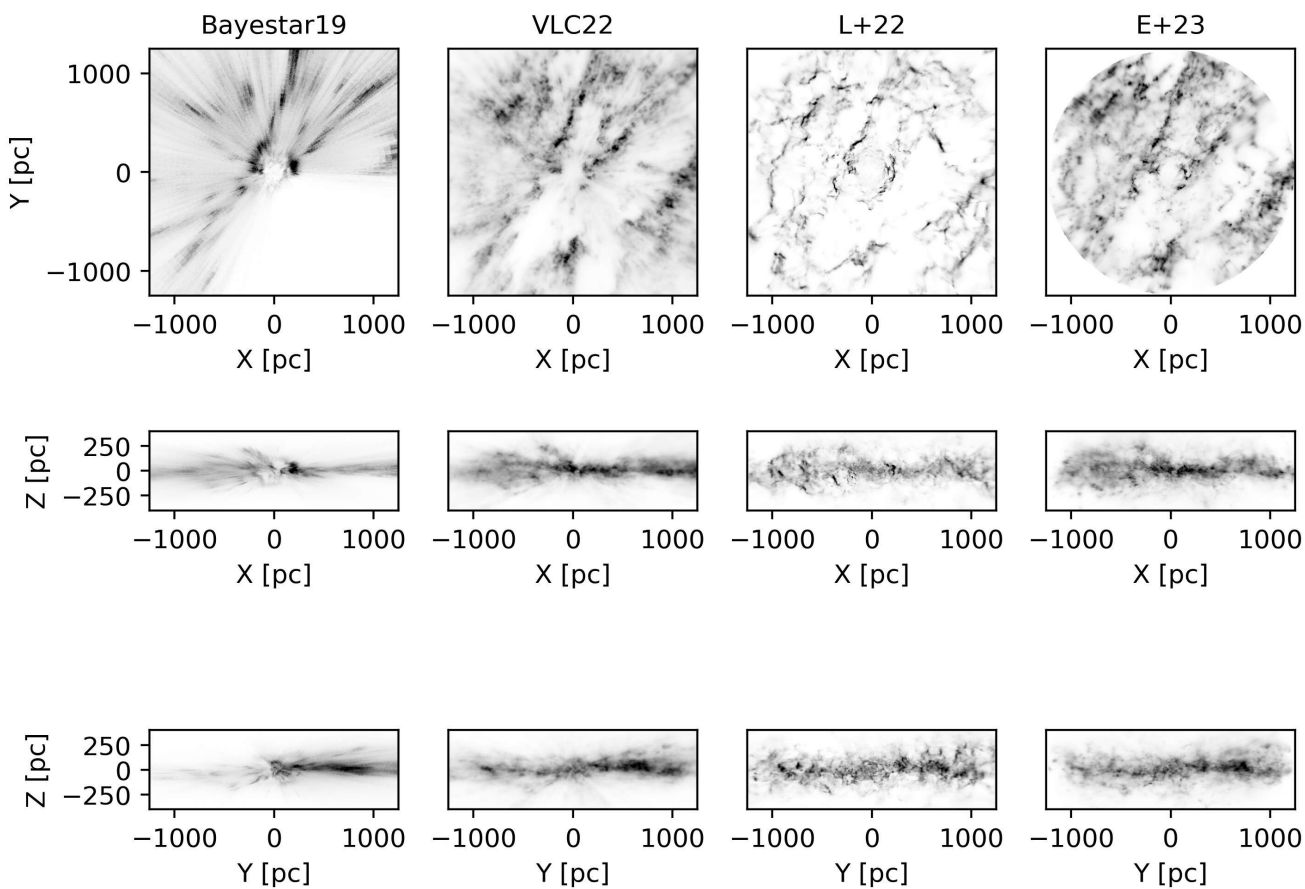

   \centering
   \includegraphics[width=0.95\hsize,keepaspectratio]{{{res/sidebyside}}}
   \caption{%
      Side by side views of the 3D dust maps from Bayestar19, VLC22, L+22, and this work, shown in heliocentric Galactic Cartesian (X, Y, Z) projections.
      The colorbars are saturated at the $99.9\%$ quantile of the respective reconstruction.
      \label{fig:sidebyside}
   }
\end{figure*}

In \Cref{fig:sidebyside}, we show 3D (X, Y, Z) projections of the maps, comparing Bayestar19, VLC22, L+22, and this work side by side.
All four maps agree on the general structure of the distribution of dust.

This work, L+22, and VLC22 have comparable distance resolutions, while Bayestar19 features comparatively few distance bins and more strongly pronounced fingers-of-god.
Compared to L+22, we feature more homogeneously extended dust clouds and significantly fewer wiggles in the distances to dust clouds.
Compared to VLC22, we feature more compact dust clouds, less grainy structures, and a higher dynamic range.
Both this work and VLC22 feature dust clouds in a comparable volume around the Sun despite the VLC22 map technically extending out farther in Galactic heliocentric X and Y.

\begin{figure*}[!htbp]
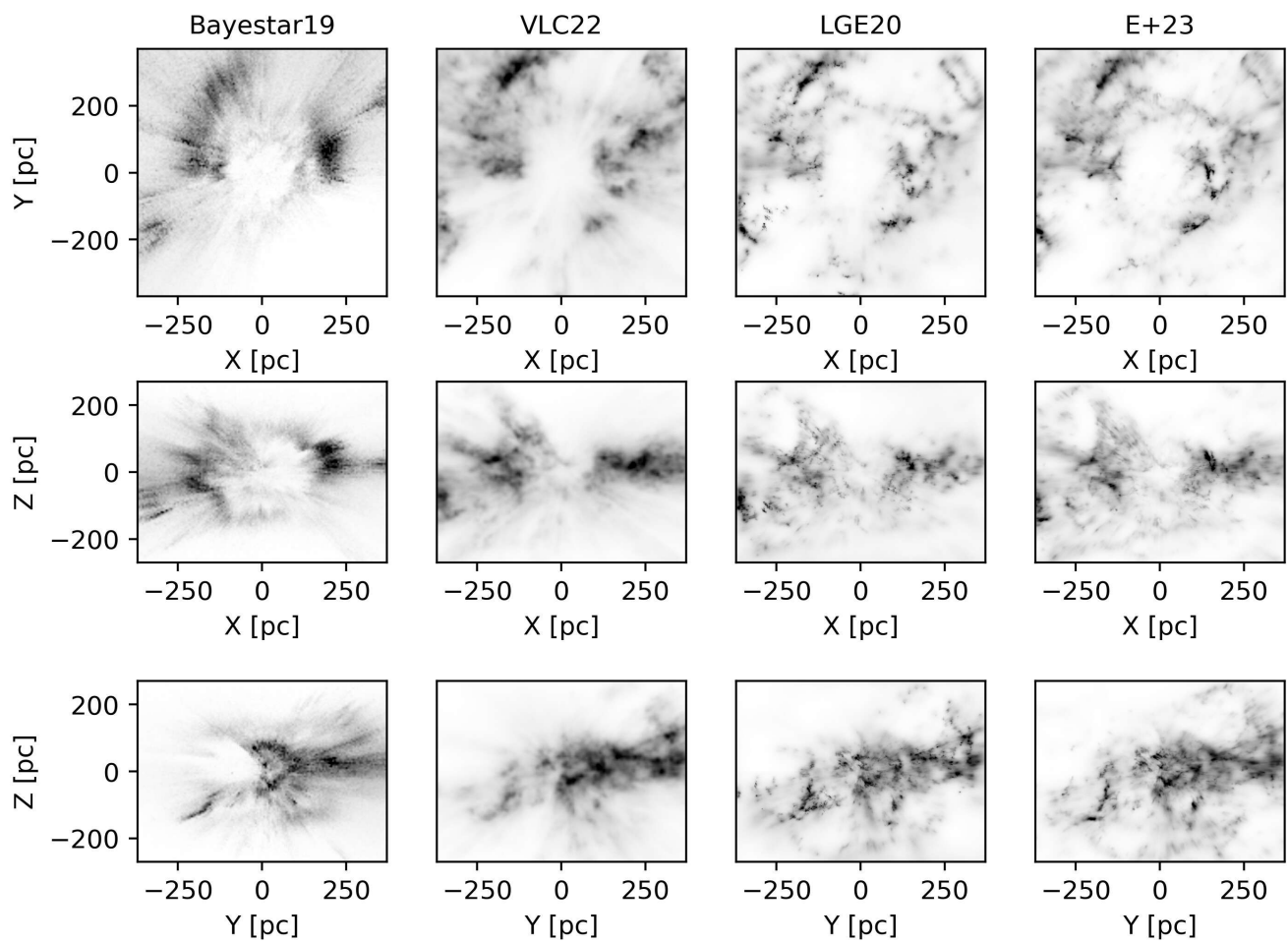

   \centering
   \includegraphics[width=0.95\hsize,keepaspectratio]{{{res/sidebyside_leike2020box}}}
   \caption{%
      Zoomed-in version of \Cref{fig:sidebyside} for the volume reconstructed in~\citet{Leike2020}, now also showing the LGE20 reconstruction for comparison.
      We omit L+22 from the comparison because the authors explicitly focus on larger volumes and trade strongly pronounced artifacts in the inner couple hundred parsecs for a larger probed volume.
      The colorbars are again saturated at the $99.9\%$ quantile of the respective reconstruction.
      \label{fig:sidebyside_leike2020box}
   }
\end{figure*}

\Cref{fig:sidebyside_leike2020box} shows the same projections for the volume reconstructed in the LGE20 map and includes the LGE20 map for comparison.
The zoom-in highlights the close similarity between this work and the LGE20 map.
All larger structures have direct correspondences in the other map, yet the distances to the structures are slightly different.
Furthermore, the LGE20 map appears slightly sharper.
The model in LGE20 is very similar to ours but uses fewer approximations.
LGE20 also uses compiled data \citep[StarHorse DR2; see][]{Anders2019}.
More work is needed to assess the validity of the sharper features in LGE20 not present in this work.
The VLC22 map is in good agreement as well but lower resolution.
Bayestar19 poorly resolves distances at the scale of the LGE20 map.

\begin{figure}[!htbp]
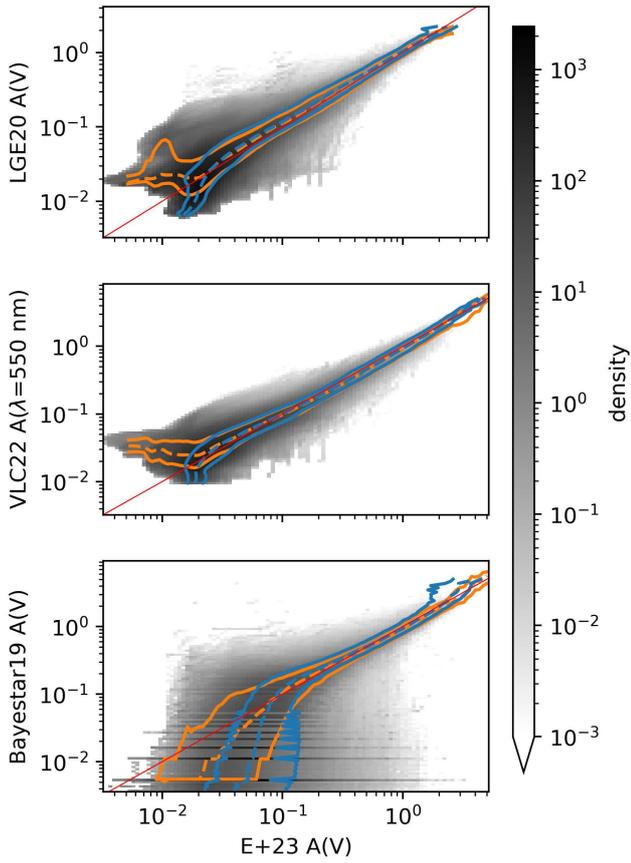

   \centering
   \includegraphics[width=0.95\hsize,keepaspectratio]{{{res/los_versus_our}}}
   \caption{%
      Histogram of the mean posterior extinction of our map versus LGE20, VLC22, and Bayestar19 for 58 million test points.
      For each pixel center of a HEALPix sphere with $N_\text{side}=64$, $1182$ test points are placed at \SI{1}{pc} intervals in distance starting at \mbox{\SI{69}{pc}.}
      The orange lines show the 16th, 50th, and 84th quantiles of the predicted extinction by LGE20, VLC22, and Bayestar19, respectively, for each bin of our mean extinction.
      The respective quantiles of our predictions in bins of the other reconstruction are shown as blue lines.
      The bisectors are shown in red.
      The colorbars are logarithmic and truncated at the lower end at \SI{1}{mmag}.
      \label{fig:los_versus_our}
   }
\end{figure}

\Cref{fig:los_versus_our} quantitatively compares our map with LGE20, VLC22, and Bayestar19.
We compared the integrated extinction for each pixel center of a HEALPix sphere with $N_\text{side}=64$ at $1182$ test points that are placed at \SI{1}{pc} intervals in distance starting at \SI{69}{pc}.
We compared integrated extinctions because slight shifts in the distances to extincted regions can be better distinguished from discrepant predictions with integrated extinctions than by directly comparing differential extinctions.
Above \SI{15}{mmag}, VLC22 and LGE20 are in good agreement with our map.
Below an $A_V$ of \SI{40}{mmag} and above an $A_V$ of approximately $\SI{3.5}{mag}$, Bayestar19 significantly deviates from our predictions.

\begin{figure*}[!htbp]
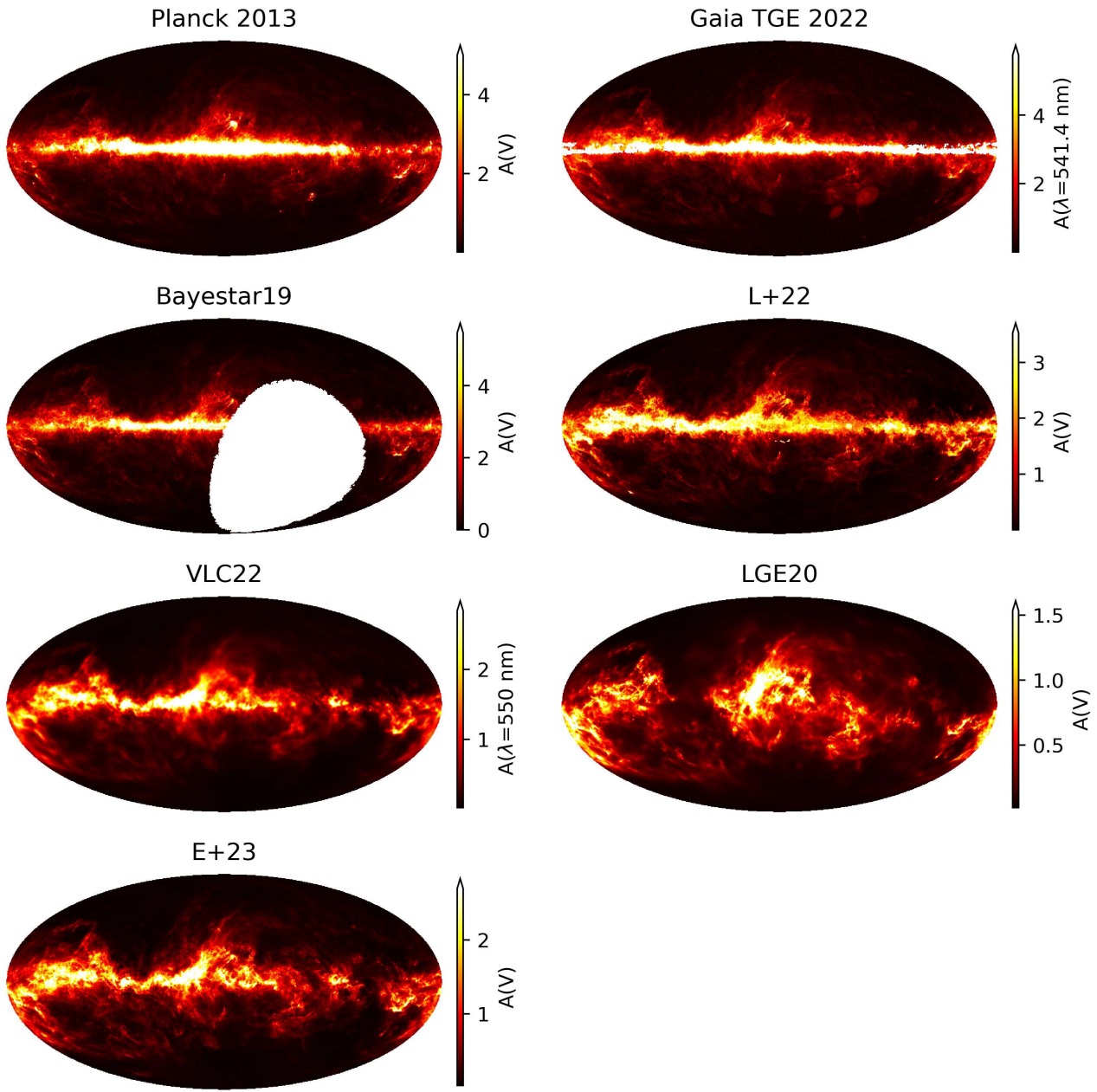

   \centering
   \includegraphics[width=0.95\hsize,keepaspectratio]{{{res/mollview_versus}}}
   \caption{%
      Mollweide projections of total integrated extinction and 3D extinction maps integrated out the maximum distance of the respective map.
      Bayestar19 reconstructs up to a maximum distance of \SI{63}{kpc} (maximum reliable distance \SI{10}{kpc}) and is integrated out to that volume.
      L+22 reconstructs up to a maximum distance of \SI{16}{kpc} but the authors trust their map only out to \SI{4}{kpc} and we integrate their map only to \SI{4}{kpc}.
      VLC22 reconstructs a heliocentric box of size $\SI{3}{kpc} \times \SI{3}{kpc} \times \SI{800}{pc}$ with $\SI[parse-numbers=false]{10^3}{pc^{3}}$ voxels with at most $<\SI{2.16}{kpc}$ in distance and is integrated out to the end of the box.
      Likewise, LGE20 reconstructs a heliocentric box of size $\rm |X| < \SI{370}{pc}, |Y| < \SI{370}{pc}, |Z| < \SI{270}{pc}$ covering at most $<\SI{590}{pc}$ in distance and is integrated out to the end of the box.
      The colorbars saturate at the respective $99\%$ quantile of the map except for the colorbar of \textit{Planck} 2013, which saturates at \SI{5}{mag} for better comparability.
      \label{fig:mollview_versus}
   }
\end{figure*}

In \Cref{fig:mollview_versus} we compare the POS view of Bayestar19, L+22, VLC22, LGE20, and this work.
The respective POS views are integrated out to the maximum distance probed by each map --- $<\SI{63}{kpc}$ in distance for Bayestar19 (maximum reliable distance \SI{10}{kpc}), $<\SI{4}{kpc}$ in distance for L+22 (authors trust structures up to \SI{4}{kpc} though the map extends to \SI{16}{kpc}), a heliocentric box of size $\SI{3}{kpc} \times \SI{3}{kpc} \times \SI{800}{pc}$ with $\SI[parse-numbers=false]{10^3}{pc^{3}}$ voxels for VLC22 with at most $\SI{2.16}{kpc}$ in distance, and a heliocentric box of size $\rm |X|, |Y| < \SI{370}{pc}, |Z| < \SI{270}{pc}$ and up to $\SI{590}{pc}$ in distance for LGE20.
In addition, we show the \textit{Planck}~2013 extragalactic dust map \citep{Planck2013} and the \textit{Gaia} total galactic extinction (TGE) 2022 map \citep{Delchambre2022}.

All maps agree on fine structures at high galactic latitudes but differ in the Galactic plane due to the difference in distance up to which the respective reconstruction extends.
3D dust reconstructions do not probe deep enough into the Galactic plane to fully recover the \textit{Planck}~2013 extragalactic dust map.
Bayestar19 and L+22 probe much deeper than VLC22, LGE20, and this work, yet they do not probe the full column of dust seen in \textit{Planck}~2013 and \textit{Gaia} TGE~2022.
Both VLC22 and our map probe up to a similar depth while LGE20 only probes dust at much closer distances.

\begin{figure*}[!htbp]
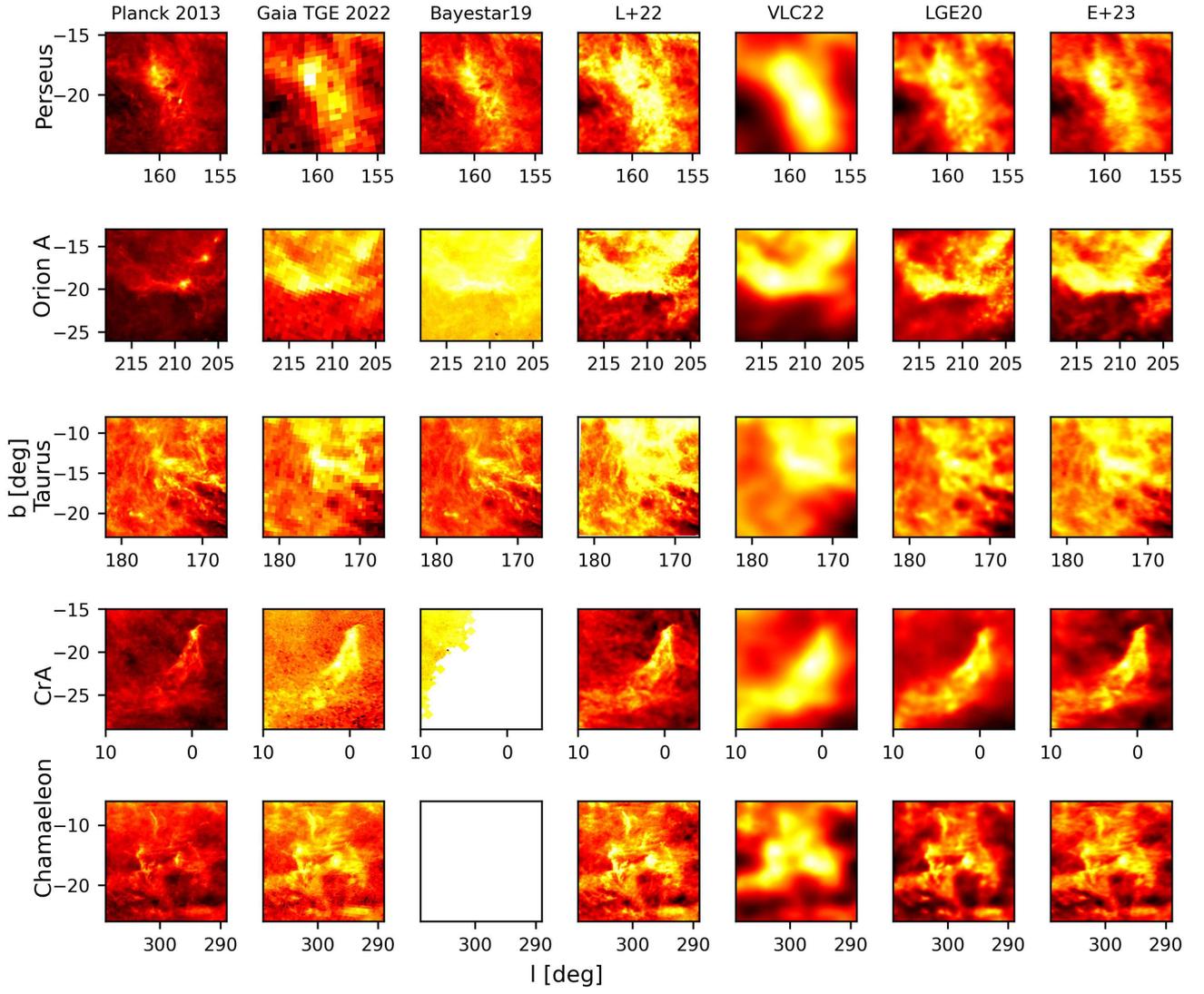

   \centering
   \includegraphics[width=0.95\hsize,keepaspectratio]{{{res/regions_of_interest}}}
   \caption{%
      Zoomed-in views toward the individual molecular clouds (Perseus, Orion, Taurus, Corona Australis, and Chamaeleon) seen in \Cref{fig:mollview_versus}.
      The colorbars are logarithmic and span the full dynamic range of the selected POS slice in every image.
      Each row is a separate region and each column a separate reconstruction.
      \label{fig:regions_of_interest}
   }
\end{figure*}

\Cref{fig:regions_of_interest} shows a zoomed-in comparison of the Perseus, Orion A, Taurus, Corona Australis (CrA), and Chamaeleon molecular clouds, integrated out to the maximum distance of each map (\SI{4}{kpc} for L+22).
Among the 3D dust reconstructions, Bayestar19 and L+22 have arguably the highest angular resolution (angular discretization of $N_\text{side}=1024$ or 3.4' and 1.9', respectively).
They resolve the high latitude dust clouds with great detail although L+22 suffers from localized artifacts in patches of the sky.
Both LGE20 (\SI{1}{pc^{3}} boxes) and this work ($N_\text{side}=256$) achieve a comparable angular resolution.
The VLC22 reconstruction (\SI[parse-numbers=false]{10^3}{pc^{3}} voxels) is noticeably lower in resolution and does not resolve the cloud substructure on the POS.
A comparison between the same molecular clouds in different distance slices in VLC22, LGE20, and our map is provided in \Cref{appx:molecular_clouds_by_distance}.

\section{Conclusions}
\label{sec:conclusions}

We present a 3D dust map with a POS and LOS resolution comparable to \citet{Leike2020} that extends out to \SI{1.25}{kpc}.
We used the distance and extinction estimates of \citet{Zhang2023}, which have much lower extinction uncertainties than competing catalogs while probing a similar number of stars.
Our reconstruction has a resolution comparable to the \SI{2}{pc} resolution of \citet{Leike2020}.
Specifically, it has an angular resolution of up to $14'$ and parsec-scale distance resolution.
Our map is in good agreement with existing 3D dust maps and improves upon them in terms of volume covered at a high spatial resolution.
The map is publicly available online\footnotemark{} and can be queried via the \texttt{dustmaps} Python package.
We anticipate that the map will be useful for a wide range of applications in studying the distribution of dust and the interstellar medium more broadly.
\footnotetext{
	\url{https://doi.org/10.5281/zenodo.8187942}
}

\begin{acknowledgements}
   We thank Joao~Alves for many fruitful discussions at the ``Self-Organization Across Scales: From nm to parsec (SOcraSCALES)'' workshop at the Munich Institute for Astro-, Particle and BioPhysics, an institute of the Excellence Cluster ORIGINS in 2022 and afterwards.
   Furthermore, we thank Jakob Roth for many invaluable discussions about the model and for providing feedback on the early versions of the reconstruction.
   We thank Alyssa Goodman for providing invaluable feedback on the late versions of the reconstructions.
   We also thank Michael A. Kuhn for providing us with a unified catalog of YSOs.
   Gordian Edenhofer acknowledges the support of the German Academic Scholarship Foundation in the form of a PhD scholarship (``Promotionsstipendium der Studienstiftung des Deutschen Volkes'').
   Catherine Zucker acknowledges that support for this work was provided by NASA through the NASA Hubble Fellowship grant \#HST-HF2-51498.001 awarded by the Space Telescope Science Institute (STScI), which is operated by the Association of Universities for Research in Astronomy, Inc., for NASA, under contract NAS5-26555.
   Philipp Frank acknowledges funding through the German Federal Ministry of Education and Research for the project ErUM-IFT: Informationsfeldtheorie für Experimente an Großforschungsanlagen (Förderkennzeichen: 05D23EO1).
   Andrew~K.~Saydjari acknowledges support by a National Science Foundation Graduate Research Fellowship (DGE-1745303).
   Andrew~K.~Saydjari and Douglas~Finkbeiner acknowledge support by NASA ADAP grant 80NSSC21K0634 ``Knitting Together the Milky Way: An Integrated Model of the Galaxy's Stars, Gas, and Dust''.
   This work was supported by the National Science Foundation under Cooperative Agreement PHY-2019786 (The NSF AI Institute for Artificial Intelligence and Fundamental Interactions).
   A portion of this work was enabled by the FASRC Cannon cluster supported by the FAS Division of Science Research Computing Group at Harvard University.
   We acknowledge support by the Max-Planck Computing and Data Facility (MPCDF).
   This work has made use of data from the European Space Agency (ESA) mission \textit{Gaia} (\url{https://www.cosmos.esa.int/gaia}), processed by the \textit{Gaia} Data Processing and Analysis Consortium (DPAC, \url{https://www.cosmos.esa.int/web/gaia/dpac/consortium}).
   Funding for the DPAC has been provided by national institutions, in particular the institutions participating in the \textit{Gaia} Multilateral Agreement.
\end{acknowledgements}

\FloatBarrier  

\bibliographystyle{aa}  
\bibliography{literature}

\begin{appendix}

   \FloatBarrier
   \section{ZGR23 in dust-free regions}
   \label{appx:zgr23_in_dust_free_regions}
   \FloatBarrier

   To gauge the reliability of the ZGR23 catalog, we analyzed the extinction to stars in dust-free regions (cf.~\citealt{Leike2019,Leike2020}).
   In dust-free regions, we would expect the extinction to be zero within the uncertainties of the catalog.
   To classify a region as dust-free, we used the \textit{Planck} dust emission map \citep{Planck2013}.
   A region is said to be dust-free if the \textit{Planck} $E(B-V)$ map is below or equal to \SI{9.5}{mmag} or approximately \SI{29}{mmag} in terms of $A_V$.

   \begin{figure}[!htbp]
      \subcaptionbox{}{%
         \includegraphics[width=0.95\hsize]{{{res/zgr23_dust_free_offset}}}
      }
      \subcaptionbox{}{%
         \includegraphics[width=0.95\hsize]{{{res/zgr23_dust_free_spread}}}
      }
      \caption{%
         Absolute and relative ZGR23 extinction in dust-free regions.
         Panel (a): Histogram of the ZGR23 extinctions in dust-free regions translated to $A_V$.
         The mean extinction in dust-free regions based on \textit{Planck} is shown as a vertical green line, and the cutoff value translated to $A_V$ for our definition of dust-free is shown in red.
         Panel (b): Same as panel (a) but the extinctions are scaled by their accompanying uncertainties.
         A truncated standard normal distribution and a standard normal distribution are plotted on top.
         Both ordinates are logarithmic.
         \label{fig:zgr23_offset}
      }
   \end{figure}

   The first panel of \Cref{fig:zgr23_offset} shows the histogram of ZGR23 extinction to stars with \texttt{quality\_flags$<$8} in dust-free regions translated to $A_V$.
   We see that the histogram of the extinction peaks at the cutoff value and coincides with the mean total extinction as measured by \textit{Planck} in those regions translated to $A_V$.
   The density of extinction values falls of exponentially after the cutoff value.
   Overall, the ZGR23 extinction seems to be in good agreement with \citet{Planck2013} for dust-free regions.

   The second panel of \Cref{fig:zgr23_offset} shows the extinction divided by their uncertainties for the stars from \Cref{fig:zgr23_offset}.
   The standardized extinctions are centered around unity, indicating that the ZGR23 extinction indeed are offset from zero by about one standard deviation in dust-free regions.
   This is in agreement with the previous finding that the extinctions are centered around the cutoff value instead of clustering around zero.
   The width around the center is comparable to a truncated standard normal distribution or a normal distribution.
   In total, about $1\%$ of the probability mass lies outside the possible range of all ZGR23 extinction with \lstinline|quality_flags<8| if we assume a normal distribution for the extinctions.

   Except for outliers far from the center, which can be captured by an outlier model, the ZGR23 catalog seems to be in agreement with POS measurements in dust-free regions and the spread around the cutoff value approximately follows a (truncated) normal distribution.
   We deem the ZGR23 catalog to be reliable for our purposes and approximate the uncertainties using a normal distribution.
   We accepted the mismodeling of a small fraction of probability mass for a simpler model (see \Cref{sec:caveats,sec:posterior_inference}).

   \FloatBarrier
   \section{Metric Gaussian variational inference}
   \label{appx:mgvi}
   \FloatBarrier

   The variational inference method MGVI approximates the true posterior $P(\xi\,|\,d)$ with a standard normal distribution in a linearly transformed space in which the posterior more closely resembles a standard normal.
   Let $Q_{\bar{\xi}}(y\,|\,d)=\mathcal{G}(y(\xi)\,|\,y(\bar{\xi}),\mathbb{1})\left|\frac{\mathrm{d}y}{\mathrm{d}\xi}\right|$ be the approximate posterior and $y(\xi): \xi \mapsto y(\xi)$ the coordinate transformation.
   In this space the transformed posterior reads $P(\xi(y)\,|\,d)\left|\frac{\mathrm{d}\xi}{\mathrm{d}y}\right|$.  
   We denote the metric of the space in which $P(\xi(y)\,|\,d)\left|\frac{\mathrm{d}\xi}{\mathrm{d}y}\right|$ is ``more'' standard normal by $M \coloneqq \frac{\mathrm{d}y}{\mathrm{d}\xi} \left(\frac{\mathrm{d}y}{\mathrm{d}\xi}\right)^\dagger$.
   Assuming $y(\xi): \xi \mapsto y(\xi)$ is known, the difficulty lies solely in finding the optimal $\bar{\xi}$ for $Q_{\bar{\xi}}$.

   Based on the Fisher information metric and Frequentist statistics, \citeauthor{Knollmueller2019} derive a coordinate transformation $y_{\bar{\xi}}(\xi)$ centered on $\bar{\xi}$ that is linear in $\xi$.
   In \citet{Frank2021}, the authors find that a set of Riemannian normal coordinates $y_{\bar{\xi}}(\xi)$ centered on $\bar{\xi}$ are an improved, nonlinear estimate of the coordinate transform $y(\xi) \approx y_{\bar{\xi}}(\xi)$.
   The improvements though come at slightly higher computational costs.
   We refer the reader to \citet{Frank2021} and \citet{Frank2022} for further details on geoVI and its relation to MGVI, the choice of metric, and an analysis of its failure modes.
   For computational reasons, we used MGVI for our inference.

   MGVI and geoVI start at a random initial position for $\bar{\xi}$ and draw $n_\text{samples}$ standard normal samples in the space of $y$.
   Next, they transform the samples to the space of $\xi$ via $y_{\bar{\xi}}$, the local, linear (respectively, nonlinear for geoVI) approximation to $y(\xi)$ at $\bar{\xi}$.
   We denote the samples in parameter space by $\{\xi_1, \dots, \xi_{n_\text{sampels}}\}$.
   Relative to the expansion point $\bar{\xi}$ the samples read $\{\Delta\xi_1\coloneqq\xi_1-\bar{\xi}, \dots, \Delta\xi_{n_\text{samples}}\coloneqq\xi_{n_\text{sampels}}-\bar{\xi}\}$.
   The samples $\{\Delta\xi_1,\dots,\Delta\xi_{n_\text{samples}}\}$ around $\bar{\xi}$ provide an empirical, sampled approximation to $Q_{\bar{\xi}}$, which we denote by $\tilde{Q}_{\bar{\xi}}$.
   MGVI and geoVI then optimize $\bar{\xi}$ of the sampled distribution, $\tilde{Q}_{\bar{\xi}}$, by minimizing the variational Kullback-Leibler (KL) divergence between $\tilde{Q}_{\bar{\xi}}$ and the true distribution $P$
   \begin{align}
      \bar{\xi}' = & \argmin_{\bar{\xi}} \mathrm{KL}\left(\tilde{Q}_{\bar{\xi}} ,\, P(\xi\,|\,d)\right)
      \\ &= \argmin_{\bar{\xi}} {\left\langle \ln{\frac{\tilde{Q}_{\bar{\xi}}}{P(\xi\,|\,d)}} \right\rangle}_{\tilde{Q}_{\bar{\xi}}}\label{eq:evi_true_kl}
      \\ &= \argmin_{\bar{\xi}} {\left\langle -\ln{P(\xi\,|\,d)} \right\rangle}_{\tilde{Q}_{\bar{\xi}}}\label{eq:evi_kl_wo_approx_dist_vol}
      \\ &= \argmin_{\bar{\xi}} \frac{-1}{n_\text{samples}} \sum_{i=1}^{n_\text{samples}} \ln{P(\Delta\xi_i - \bar{\xi}\,|\,d)}\label{eq:evi_kl_sampled}
      \ .
   \end{align}
   They keep the relative samples $\{\Delta\xi_1\coloneqq\xi_1-\bar{\xi}, \dots, \Delta\xi_{n_\text{samples}}\coloneqq\xi_{n_\text{sampels}}-\bar{\xi}\}$ fixed during the optimization and only vary $\bar{\xi}$.
   Finally, they update the expansion point $\bar{\xi}$ to the new found optimum $\bar{\xi}'$.

   After the minimization, MGVI and geoVI draw a new set of samples, transform them via a local (linear) expansion of $y$, and then minimize again.
   The drawing of samples and minimization is repeated until a fixed point for $\bar{\xi}$ is reached.
   \Cref{alg:expansion_point_vi} summarizes the algorithmic steps of the variational approximation to the true posterior.

   \begin{algorithm}
      \SetAlgoLined
      \DontPrintSemicolon

      \SetKwProg{Def}{def}{:}{}
      \Def{sample($\xi$, $n_\text{samples}$)}{
         $\dots$\;
         \Return $\{\xi_1, \dots, \xi_{n_\text{sampels}}\}$\;
      }

      $\bar{\xi}$ $\gets$ \dots  \tcp{random init}\;
      \While{$\bar{\xi}$ $\text{not converged}$}{
      $\{\xi_1, \dots, \xi_{n_\text{sampels}}\} \gets$ sample($\bar{\xi}$, $n_\text{samples}$)\;
      $\{\Delta\xi_1, \dots, \Delta\xi_{n_\text{samples}}\} \gets \{\xi_1-\bar{\xi}, \ldots, \xi_{n_\text{sampels}}-\bar{\xi} \}$\;
      $\bar{\xi} \gets \argmin_{\bar{\xi}} \frac{-1}{n_\text{samples}} \sum_{i=1}^{n_\text{samples}} \ln{P(\Delta\xi_i + \bar{\xi}\,|\,d)}$\;
      }

      \caption{%
         Pseudocode for the MGVI and geoVI expansion point variational inference scheme.
         \label{alg:expansion_point_vi}
      }
   \end{algorithm}

   \FloatBarrier
   \section{Extinction within the innermost \SI{69}{pc}}
   \label{appx:extinction_within_the_innermost_x_pc}
   \FloatBarrier

   By construction of our likelihood, we know that our model is biased high for low extinction values because we neglected the positivity prior of the ZGR23 catalog.
   This imprints on to the map in the form of a thin layer of dust extinction right at the beginning of the modeled volume.
   As voxels in our reconstruction are correlated, the extincted first layer of voxels pulls the next layer of voxels to slightly higher extinctions too.
   At \SI{69}{pc} the differential extinction as a function of distance reaches a local minimum, and we expect little to now influence of the innermost layers of voxels.
   Thus, we determined \SI{69}{pc} to be our cutoff.

\begin{figure}[!htbp]
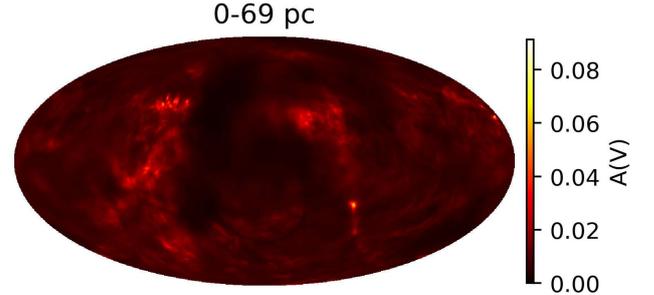

      \centering
      \includegraphics[width=0.95\hsize,keepaspectratio]{{{res/zeroth_mollview}}}
      \caption{%
         Mollweide projection of the integrated $A_V$ extinction in the innermost \SI{69}{pc}, which is likely dominated by spurious effects and is therefore excluded from the reconstruction.
         The colorbar is linear and covers the full range of the extinction that is cut out of the reconstruction.
         \label{fig:zeroth_mollview}
      }
   \end{figure}

   \Cref{fig:zeroth_mollview} shows the integrated extinction that is cut out from the final reconstruction.
   Most of the extinction is likely spurious.
   Overall, no structure contributes a significant amount of extinction.
   However, to be consistent with ZGR23, we suggest adding the removed extinction back to the map when comparing integrated extinction.

   \FloatBarrier
   \section{Extinction catalog}
   \label{appx:extinction_catalog}
   \FloatBarrier

   We release a catalog of expected extinction for all stars within the subset of the ZGR23 catalog that we used for our reconstruction (see~\Cref{sec:data}).
   We predict the expected extinction conditional on the known parallax including parallax uncertainties.
   Our prediction is the best guess of our model for the extinction toward a star but is not necessarily the best guess for the extinction at the mean parallax of the star.

   Our extinction predictions (see \Cref{sec:priors,sec:likelihood:response}) differ from the extinctions in the ZGR23 catalog by coupling the individual stars via the 3D dust extinction density.
   By virtue of every star depending on all nearby stars via the prior, our extinction predictions come in the form of joint predictions for all stars.
   In regions where the 3D dust extinction density is well constrained, the joint predictions to first order factorize into predictions for individual stars, and we can compute expected extinctions for individual stars and their uncertainties.

   Our catalog of extinction includes the innermost \SI{69}{pc} from the beginning of our grid and the outer \SI{550}{pc} beyond \SI{1.25}{kpc} that we cut away in the 3D map.
   We advise caution when analyzing the stars of our catalog within those regions as they might carry additional biases.
   Details on why these regions were removed from the final map are given in \Cref{sec:caveats,sec:posterior_inference}.

   \begin{figure}[!htbp]
      \subcaptionbox{}{%
         \includegraphics[width=0.95\hsize]{{{res/zgr23_over_ours}}}
      }
      \subcaptionbox{}{%
         \includegraphics[width=0.95\hsize]{{{res/zgr23_unc_over_ours}}}
      }
      \caption{\label{fig:zgr23_versus_ours}%
         ZGR23's extinction versus our predicted extinction for Gaia BP/RP stars.
         Panel (a): Our mean posterior extinctions versus the ZGR23 extinctions to stars as 2D histogram.
         The $16^\text{th}$, $50^\text{th}$, and $84^\text{th}$ quantiles of the ZGR23 extinctions for each bin of our mean extinction are shown as blue lines.
         The respective quantiles of our predictions in bins of the ZGR23 extinctions are shown as orange lines.
         Panel (b): Same comparison but for our posterior mean predictions for the ZGR23 measurement uncertainties, $\sqrt{\left[n_\sigma(\xi) \cdot \sigma_A\right]^2+\sigma_a^2}$, versus the ZGR23 uncertainties.
         Note that the predictions for the ZGR23 measurement uncertainties are not the uncertainties of our extinction predictions.
         See \Cref{sec:likelihood} and in particular \Cref{eq:total_likelihood} for further details on the quantities shown here.
         The bisectors are shown in red.
         The colorbars are logarithmic.
      }
   \end{figure}

   The top panel of \Cref{fig:zgr23_versus_ours} compares the ZGR23 extinctions to our mean extinction predictions for Gaia BP/RP stars.
   Overall, our mean extinctions are in very good agreement with the extinctions in the ZGR23 catalog for the vast majority of stars.
   However, below \SI{50}{mmag} and above \SI{4}{mag,} our extinction predictions deviate from the predictions in ZGR23.
   In any given extinction bin from ZGR23 respectively our work, we would expect half of the respective other extinctions to be below the bisector and the other half to be above.
   At \SI{50}{mmag}, 34\% more stars than expected have higher extinctions than the corresponding extinctions in ZGR23.
   The difference further widens for lower ZGR23 extinctions.
   At \SI{4}{mag}, 34\% more stars than expected have lower extinctions than the corresponding extinctions in ZGR23.

   The bottom panel of \Cref{fig:zgr23_versus_ours} shows our and the ZGR23 extinction uncertainties.
   We note that our extinction uncertainties are predictions for the measured uncertainties of the ZGR23 catalog $\left[n_\sigma(\xi) \cdot \sigma_A\right]^2+\sigma_a^2(\rho(\xi))$ and not the uncertainties of our extinction predictions $\mathrm{std}(\bar{a})$ (see \Cref{sec:likelihood}).
   Overall, both uncertainties agree well for the vast majority of stars.
   At low extinctions uncertainties our uncertainties only marginally inflate the ZGR23 uncertainties.
   However, at high extinction uncertainties, our predictions cover a larger range, and we find that the ZGR23 significantly underpredicts our extinction uncertainties of stars.

   \begin{figure}[!htbp]
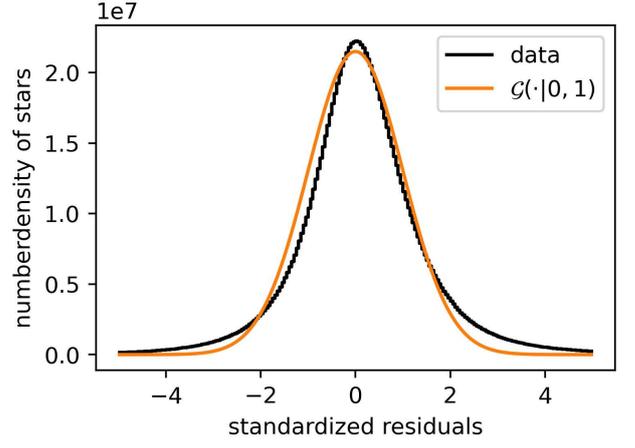

      \centering
      \includegraphics[width=0.95\hsize]{{{res/standardized_residuals}}}
      \caption{%
         Mean standardized extinctions: $(A - \bar{a}) / \sqrt{(n_\sigma \cdot \sigma_A)^2 + \sigma^2_a}$ (see \Cref{sec:likelihood} and in particular \Cref{eq:total_likelihood}) within the range $-5$ to $5$.
         \label{fig:standardized_extinction_prediction}
      }
   \end{figure}

   \Cref{fig:standardized_extinction_prediction} summarizes the extinctions and the extinction uncertainties of both ZGR23 and our predictions into a single histogram of the mean standardized extinction.
   A standard Gaussian is shown as reference.
   The mean standardized residuals have two slight overdensities at each tail compared to the Gaussian.

   \begin{figure}[!htbp]
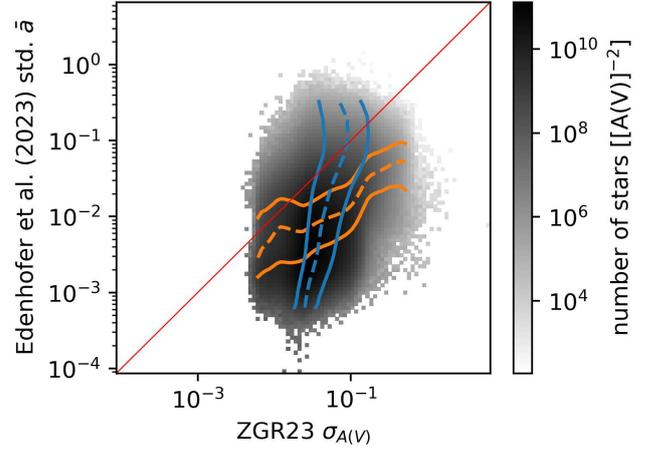

      \centering
      \includegraphics[width=0.95\hsize]{{{res/zgr23_unc_over_our_std_of_mean}}}
      \caption{\label{fig:actual_zgr23_versus_ours_unc}%
         Similar to \Cref{fig:zgr23_versus_ours} but for the posterior standard deviation of our extinctions versus the ZGR23 uncertainties.
         The $16^\text{th}$, $50^\text{th}$, and $84^\text{th}$ quantiles of the ZGR23 uncertainties for each bin of our standard deviation are shown as blue lines.
         The respective quantiles of our standard deviation in bins of the ZGR23 uncertainties are shown as orange lines.
         The bisectors are shown in red.
         The colorbars is logarithmic.
      }
   \end{figure}

   \Cref{fig:actual_zgr23_versus_ours_unc} shows the posterior standard deviation of our extinction predictions versus the ZGR23 uncertainties.
   Our model yields approximately one order in magnitude lower extinction uncertainties than the ZGR23 uncertainties for the vast majority of stars.
   The effect is less pronounced for low ZGR23 extinction uncertainties.

   Our predictions for the extinction to stars theoretically contain more information since we allow for the cross-talk of nearby stars via the 3D distribution of dust and thus might be more accurate.
   However, the ZGR23 catalog might yield better results in practice because it does not discretize the 3D volume within which the stars reside.
   By discretizing the modeled volume we can produce contradicting data that in a continuous space is non-contradicting, for example by putting highly extincted stars that lie in a dust cloud into the same voxel as less extincted stars that are adjacent to the dust cloud.
   Overall, both predictions agree very well for stars below between \SI{50}{mmag} and \SI{4}{mag}.
   More work is needed to validate the discrepant predictions at very low and very high extinctions.

   \FloatBarrier
   \section{\SI{2}{kpc} reconstruction}
   \label{appx:2kpc_reconstruction}
   \FloatBarrier

   In \Cref{sec:posterior_inference} we describe how we iteratively increase the distance out to our maximum reconstructed distance.
   We did so to improve the convergence of the reconstruction.
   We also tried naively reconstructing the full volume at once.
   Using all the available data is computationally prohibitive, so we limited the reconstruction to high quality data using \lstinline|quality_flags==0|, $\sigma_A \leq 0.04$, and $\nicefrac{\sigma_\omega}{\omega} < 0.33$.

   We used $\nicefrac{1}{(\omega-\sigma_\omega)}<\SI{3}{kpc}$ and $\nicefrac{1}{(\omega+\sigma_\omega)}>\SI{40}{pc}$ to select the stars within a \SI{3}{kpc} sphere.
   To further speed up the inference, we started the inference using at first only a sample of $10\%$, then $20\%$, $45\%$, $67\%,$ and finally $100\%$ of the stars.
   In total, we selected $\num[group-separator={,}]{59334214}$ stars.
   After the inference, we cut away the outermost \SI{1}{kpc} of the sphere of the data-constrained region to avoid degradation effects due to the thinning out of stars at the edge.
   The overall reconstructed volume after removing the outermost HEALPix spheres extends out to \SI{2}{kpc} in distance.

   \begin{figure*}[!htbp]
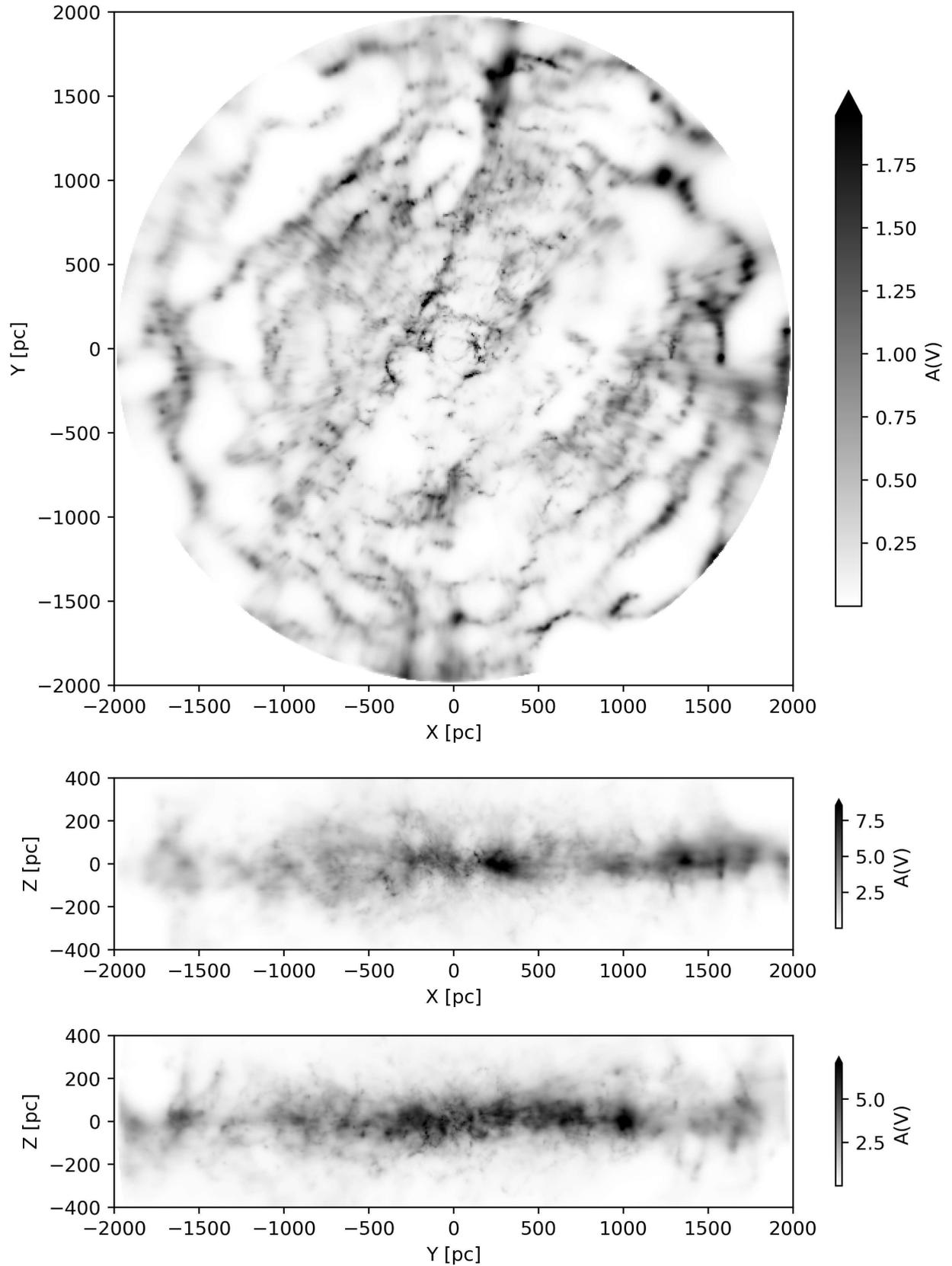

      \centering
      \includegraphics[width=0.9\hsize]{{{res/galactic_2kpc}}}
      \caption{%
         Axis parallel projections of the reconstructed dust extinction in a box of dimensions $\SI{4}{kpc} \times \SI{4}{kpc} \times \SI{0.8}{kpc}$ centered on the Sun.
         The colorbar is linear and saturates at the $99.9\%$ quantile.
         \label{fig:galactic_2kpc}
      }
   \end{figure*}

   \begin{figure*}[!htbp]
      \centering
      \includegraphics[width=0.9\hsize]{{{res/galactic_2kpc_with_ysos}}}
      \caption{%
         Same as \Cref{fig:galactic_2kpc} but with a catalog of clusters of YSOs~\citep{Kuhn2023YSO} based on \citet{Kuhn2021}, \citet{Winston2020}, and \citet{Marton2022} shown a blue dots on top of the reconstruction; their distance uncertainties are shown as extended lines.
         \label{fig:galactic_2kpc_with_ysos}
      }
   \end{figure*}

   The reconstruction is shown in \Cref{fig:galactic_2kpc} and again in \Cref{fig:galactic_2kpc_with_ysos} with a catalog of YSO clusters~\citep{Kuhn2023YSO} overlaid on top.
   It shows the same large-scale features as the smaller reconstruction discussed in the main text.
   The distribution of dense dust clouds is in agreement with the positions of YSO clusters within the distance uncertainties of the YSO clusters.
   Compared to \Cref{fig:galactic} the reconstruction is less detailed and features more pronounced artifacts.

   We used the larger reconstruction to validate the inference of the smaller one.
   Specifically, we used the larger reconstruction to ensure that structures aligned with or close to the radial boundaries at which we increased the distance of the main reconstruction are independent of the locations at which we increased the distance covered.

   We release the larger reconstruction as an additional data product together with the main reconstruction.
   We advise using the main reconstruction for all regions that fall within its volume.
   Care should be taken when interpreting small-scale features or structures at high distances in the larger reconstruction.

   \FloatBarrier
   \section{Using the reconstruction}
   \label{appx:using_the_reconstruction}
   \FloatBarrier

   All data products are made publicly available online\footnotemark{}.
   The data products are stored in the FITS file format.
   The main data products are the posterior samples of the spatial 3D distribution of dust extinction discretized to HEALPix spheres at logarithmically spaced distances.
   For convenience, we also provide the posterior mean and standard deviation of the samples of the HEALPix spheres at logarithmically spaced distances.
   \footnotetext{
		\url{https://doi.org/10.5281/zenodo.8187942}
   }

   We additionally interpolated the posterior mean and standard deviation to a Cartesian grid.
   The interpolation was carried out at a lower discretization using $\SI[parse-numbers=false]{2^3}{pc^{3}}$ voxels to keep the file size reasonably small.
   We recommend re-interpolating the map at a higher discretization for the study of individual regions within the map.

   We release the interpolation script as part of the data release.
   Its signature reads \lstinline|interp2box.py [-h] [-o OUTPUT_DIRECTORY] [-b BOX] healpix_path|.
   A box is a string of two tuples separated by two colons.
   The first tuple specifies the number of voxels along each axis of the box and the second tuple specifies the corners of the box in parsecs in heliocentric coordinates.
   To interpolate the map to a box with $1051 \times 1051 \times 351$ voxels of size $\rm |X|, |Y| \leq \SI{2100}{pc}$ and $\rm |Z| \leq \SI{700}{pc}$, use
   \lstinline|interp2box.py -b '(1051,1051,351)::((-2100,2100),(-2100,2100),(-700,700))' -- mean_and_std_healpix.fits|.

   In addition, we interpolated the posterior mean and standard deviation to galactic longitude, latitude, and distance.
   The signature of the interpolation script reads \lstinline|interp2lbd.py [-h] [-o OUTPUT_DIRECTORY] [-b BOX] healpix_path|.
   Its behavior is similar to \lstinline|interp2box.py| but the box is specified in terms of galactic longitude, latitude, and distance in units of degrees, degrees, and parsecs, respectively.

   Both scripts require the Python packages numpy \citep{Harris2020}, astropy \citep{Astropy2013,Astropy2018,Astropy2022}, and \texttt{healpy} \citep{Gorski2005,Zonca2019}.
   Depending on the number of output voxels, the interpolation can be very memory intensive and computationally expensive.

   \FloatBarrier
   \section{Molecular clouds by distance}
   \label{appx:molecular_clouds_by_distance}
   \FloatBarrier

   \Cref{fig:regions_of_interest_Perseus,fig:regions_of_interest_Orion_A,fig:regions_of_interest_Taurus,fig:regions_of_interest_CrA,fig:regions_of_interest_Chamaeleon} depict zoomed-in views of \Cref{fig:mollview_versus} at different distant slices, alongside histograms comparing the extinction of VLC22, LGE20, and our mean map toward Perseus, Orion A, Taurus, Corona Australis, and Chamaeleon.
   The top panels depict the same POS views for the clouds as in \Cref{fig:regions_of_interest}, but now showing the dust extinction in finite distance bins, rather than integrated over the full distance range.
   A low-resolution 3D interactive figure of the reconstruction including all the above molecular clouds is available online\footnotemark{}.
   \footnotetext{
	   \url{https://faun.rc.fas.harvard.edu/czucker/Paper_Figures/3D_Dust_Edenhofer2023.html}
   }

   The bottom panels of the figures are akin to \Cref{fig:los_versus_our} but compare the extinction within the selected distance and POS area of the respective molecular cloud only.
   The probing points to which we integrated are spaced in the distance range of the top panel.
   We started the integration at the lowermost distance of the top panel and successively integrated out in steps of \SI{0.5}{pc}.
   The probing points are spaced ${15'}$ apart along the POS.

   LGE20 and our map agree well for Perseus, Orion A, Taurus, and Chamaeleon.
   Some structures appear at slightly larger distances in our map as indicated by the arches above the diagonal in the second panels.
   LGE20 saturates sooner than our map as it flattens off at high extinctions below the diagonal.
   The flattening off is less pronounced for Chamaeleon.

   Corona Australis is an outlier among the generally good level of agreement.
   In the lower panel, there is noticeably more density off the diagonal.
   We see arches both above and below the diagonal indicating that some structures are closer while others are farther in our map compared with LGE20.
   While LGE20 puts the head of Corona Australis in the distance slice ranging from \SIrange{140}{160}{pc}, our map puts the head of Corona Australis dozens of parsecs farther in distance.
   Possible reasons for this shift include an insufficient number of stars to constrain the distance to the head of Corona Australis or a failure mode in our posterior approximation.

   VLC22 is much lower in resolution and does not resolve degree-scale high extinction regions.
   The extinction in VLC22 within the selected distance slice is much lower than in our map and by extension in LGE20.
   The missing extinction partially lies outside the selected box as it is strongly smeared out radially.

   \begin{figure}[!htbp]
      \subcaptionbox{}{%
         \includegraphics[width=0.95\hsize]{{{res/regions_of_interest_Perseus_by_distance}}}
      }
      \subcaptionbox{}{%
         \hspace*{0.175\hsize}
         \includegraphics[width=0.7\hsize]{{{res/los_versus_our_for_regions_of_interest_Perseus}}}
      }
      \caption{\label{fig:regions_of_interest_Perseus}%
         Comparison of different dust maps for Perseus.
         Panel (a): Zoomed-in view of \Cref{fig:mollview_versus} toward Perseus akin to \Cref{fig:regions_of_interest}, in different distant slices.
         Columns depict dust reconstructions, while rows depict distance slices.
         The logarithmic colorbars are separate for the reconstructions but shared for different distance slices.
         Panel (b): Comparison of the mean posterior extinction integrated from the lowest distance of Panel (a) to regularly spaced points in the distance range of Panel (a).
         The extinction predictions for our map versus LGE20 and VLC22 are shown as histograms.
         The binning is linear and the colorbar logarithmic.
         The bisectors are shown in red.
      }
   \end{figure}

   \begin{figure}[!htbp]
      \subcaptionbox{}{%
         \includegraphics[width=0.95\hsize]{{{res/regions_of_interest_Orion_A_by_distance}}}
      }
      \subcaptionbox{.}{%
         \hspace*{0.175\hsize}
         \includegraphics[width=0.7\hsize]{{{res/los_versus_our_for_regions_of_interest_Orion_A}}}
      }
      \caption{\label{fig:regions_of_interest_Orion_A}%
         Same as \Cref{fig:regions_of_interest_Perseus} but for Orion A.
      }
   \end{figure}

   \begin{figure}[!htbp]
      \subcaptionbox{}{%
         \includegraphics[width=0.95\hsize]{{{res/regions_of_interest_Taurus_by_distance}}}
      }
      \subcaptionbox{}{%
         \hspace*{0.175\hsize}
         \includegraphics[width=0.7\hsize]{{{res/los_versus_our_for_regions_of_interest_Taurus}}}
      }
      \caption{\label{fig:regions_of_interest_Taurus}%
         Same as \Cref{fig:regions_of_interest_Perseus} but for Taurus.
      }
   \end{figure}

   \begin{figure}[!htbp]
      \subcaptionbox{.}{%
         \includegraphics[width=0.95\hsize]{{{res/regions_of_interest_CrA_by_distance}}}
      }
      \subcaptionbox{}{%
         \hspace*{0.175\hsize}
         \includegraphics[width=0.7\hsize]{{{res/los_versus_our_for_regions_of_interest_CrA}}}
      }
      \caption{\label{fig:regions_of_interest_CrA}%
         Same as \Cref{fig:regions_of_interest_Perseus} but for Corona Australis (CrA).
      }
   \end{figure}

   \begin{figure}[!htbp]
      \subcaptionbox{}{%
         \includegraphics[width=0.95\hsize]{{{res/regions_of_interest_Chamaeleon_by_distance}}}
      }
      \subcaptionbox{}{%
         \hspace*{0.175\hsize}
         \includegraphics[width=0.7\hsize]{{{res/los_versus_our_for_regions_of_interest_Chamaeleon}}}
      }
      \caption{\label{fig:regions_of_interest_Chamaeleon}%
         Same as \Cref{fig:regions_of_interest_Perseus} but for Chamaeleon.
      }
   \end{figure}

\end{appendix}

\end{document}